\documentstyle[prb,aps,epsfig,a4]{revtex}
\begin{document}        

\title{\bf Study of the process $e^+e^- \to \pi^+\pi^-\pi^0$ in the energy
           region $\sqrt[]{s}$ below 0.98 GeV.}
\author{ M.N.Achasov\thanks{e-mail: achasov@inp.nsk.su,
FAX: +7(383-2)34-21-63},
	 K.I.Beloborodov, A.V.Berdyugin, A.G.Bogdanchikov,
	 A.V.Bozhenok, A.D.Bukin, D.A.Bukin, T.V.Dimova,
	 V.P.Druzhinin, V.B.Golubev, I.A.Koop, A.A.Korol,
         S.V.Koshuba, A.P.Lysenko, E.V.Pakhtusova,
	 S.I.Serednyakov,  V.V.Shary,
	 Yu.M.Shatunov, Z.K.Silagadze, A.N.Skrinsky,
	 A.A.Valishev, A.V.Vasiljev}
\address{Budker Institute of Nuclear Physics,  \\
         Siberian Branch of the Russian Academy of Sciences \\
	 and Novosibirsk State University, \\
	 11 Lavrentyev,Novosibirsk, \\
	 630090, Russia}
\date{\today}
\maketitle

\begin{abstract}
 The cross section of the process $e^+e^-\to \pi^+\pi^-\pi^0$ was measured in
 the Spherical Neutral Detector (SND) experiment at the VEPP-2M collider in the
 energy region $\sqrt[]{s}$ below 980 MeV. This measurement was based on
 about $1.2 \times 10^6$ selected events. The obtained cross section was
 analyzed together with the SND and DM2 data in the energy region $\sqrt[]{s}$
 up to 2 GeV. The $\omega$-meson parameters: $m_\omega=782.79\pm 0.08\pm 0.09$
 MeV, $\Gamma_\omega=8.68\pm 0.04\pm 0.15$ MeV and
 $\sigma(\omega\to 3\pi)=1615\pm 9\pm 57$ nb were obtained. It was found that
 the experimental data cannot be described by a sum of only $\omega$, $\phi$,
 $\omega^\prime$ and $\omega^{\prime\prime}$ resonances contributions. This
 can be interpreted as a manifestation of $\rho\to 3\pi$ decay, suppressed by
 $G$-parity, with relative probability
 $B(\rho\to 3\pi) = (1.01\pm^{0.54}_{0.36}\pm 0.034) \times 10^{-4}$.
 
\end{abstract}

\section{Introduction}
 The cross section of $e^+e^- \to \pi^+\pi^-\pi^0$ process in the energy region
 $\sqrt[]{s}<$ 2 GeV is determined by the transitions of light vector mesons
 V ($V=\omega,\phi,\omega^\prime,\omega^{\prime\prime}$) into the final state:
 $V\to\pi^+\pi^-\pi^0$. The $V\to\pi^+\pi^-\pi^0$ branching ratios for vector
 mesons with isospin $I=0$ are large:
 $B(\omega\to 3\pi) \simeq 0.9$, $B(\phi\to 3\pi) \simeq 0.15$ \cite{pdg},
 $B(\omega^\prime\to 3\pi)\sim 1$, $B(\omega^{\prime\prime}\to 3\pi)\sim 0.5$
 \cite{pi3mhad}, and thus the $e^+e^-\to\pi^+\pi^-\pi^0$ cross section 
 measurements are important for study of these resonances.
 The $\rho\pi$ intermediate state dominates in these transitions.
 The $V\to\pi^+\pi^-\pi^0$ transition can also proceed via mechanisms 
 suppressed by 
 the $G$-parity: $V\to\omega\pi^0\to\pi^+\pi^-\pi^0$
 ~\cite{thrhoom,pi3mhad} or $V\to\rho\pi\to\pi^+\pi^-\pi^0$
 ($V=\rho,\rho^\prime,\rho^{\prime\prime}$). The studies of the
 $e^+e^- \to \pi^+\pi^-\pi^0$ reaction allow to determine the vector mesons
 parameters and provide information on the $OZI$ rule violation in 
 $\phi\to 3\pi$
 decay and on the $G$-parity violation in the processes
 $\rho,\rho^{\prime(\prime\prime)}\to 3\pi$. 

 The process $e^+e^- \to \pi^+\pi^-\pi^0$ in the energy region $\sqrt[]{s}$
 below 2200 MeV was studied in several experiments during the last 30 years.
 The $\omega$-meson production region was studied in the
 Ref.\cite{aug,benak,olyaomega,ndomega,nd,kmd,kmd2omega} and studies of the
 $\phi$-meson energy domain were reported in
 Ref.\cite{cosm,parr,parr2,buk,kmd2phi1,kmd2phi2,nd}. In
 Ref.\cite{cord,olyaphiomega,ndnerez,nd} the $e^+e^- \to \pi^+\pi^-\pi^0$ cross
 section was studied in the wide energy region $\sqrt[]{s}$ from 660 up to 1100
 MeV and in the Ref.\cite{ndrho} the upper limit was imposed on the $G$-parity
 suppressed decay $\rho\to 3\pi$. The $e^+e^- \to \pi^+\pi^-\pi^0$
 cross section measurements in the $\omega^\prime$ and $\omega^{\prime\prime}$
 resonances energy region ($\sqrt[]{s} = 1100$ -- 2200 MeV) were reported in
 Ref.\cite{m3n,mea,gg2,dm1,ndnerez,nd,dm2}. 

 Recently the process  $e^+e^-\to\pi^+\pi^-\pi^0$ was also studied with the
 Spherical Neutral Detector (SND)\cite{sndmhad,phi98,dplphi98,pi3mhad}, 
 the process dynamics was analyzed and the cross section
 was measured in the energy region $\sqrt[]{s}$ from 980 to 1380 MeV. Here we
 present the $e^+e^-\to\pi^+\pi^-\pi^0$ cross section measurement in the energy
 region $\sqrt[]{s}$ below 980 MeV. The obtained cross section was analyzed
 together with the SND \cite{phi98,pi3mhad} and DM2 \cite{dm2} data in the 
 energy region up to 2000 MeV.
 
\section{Experiment}

 The SND detector \cite{sndnim} operated from 1995 to 2000 at the
 VEPP-2M \cite{vepp2} collider in the energy range $\sqrt[]{s}$ from 360 to
 1400 MeV. The detector contains several subsystems. The tracking system
 includes two cylindrical drift chambers. The three-layer spherical
 electromagnetic calorimeter is based on NaI(Tl) crystals.
 The muon/veto system consists of plastic scintillation counters and two
 layers of streamer tubes. The calorimeter energy and angular resolutions
 depend on the photon energy as
 $\sigma_E/E(\%) = {4.2\% / \sqrt[4]{E(\mbox{GeV})}}$ and
 $\sigma_{\phi,\theta} = {0.82^\circ / \sqrt[]{E(\mathrm{GeV})}} \oplus 
 0.63^\circ$ 
 The tracking system angular resolution is about $0.5^\circ$ and $2^\circ$ for
 azimuthal and polar angles respectively.

 In 1998 -- 2000 the SND detector collected data in the energy region 
 $\sqrt[]{s}<980$
 MeV with integrated luminosity about $10.0~\mbox{pb}^{-1}$. For the
 luminosity measurements, the processes $e^+e^- \to e^+e^-$ and
 $e^+e^- \to \gamma\gamma$ were used. In this work the luminosity measured by
 using $e^+e^- \to \gamma\gamma$ was used for normalization, because in the
 $\rho$-meson energy region the contribution of the $e^+e^-\to\pi^+\pi^-$ 
 background to the $e^+e^- \to e^+e^-$ process is rather large. The systematic
 error of the integrated luminosity determination is estimated to be 2\%.
 Since luminosity
 measurements by  $e^+e^- \to e^+e^-$ and $e^+e^- \to \gamma\gamma$ reveal a
 systematic spread of about 1\%, this  was added to the statistical error of
 the luminosity determination in each energy point. The statistical accuracy
 was better than 1\%.

 The beam energy was calculated from the magnetic field value in the bending
 magnets and revolution frequency of the collider. The relative accuracy of the
 energy setting for each energy point is about 0.1 MeV, while the common shift
 of the energy scale for all points within the scan can amount to 0.5 MeV. 
 In the three energy points in the vicinity of the
 $\omega$-resonance peak the beam energy was calibrated using resonant
 depolarization method \cite{rdp}. The accuracy of center of mass energy
 calibration is 0.04 MeV. In order to correct the calculated beam energy, the
 common shifts of the energy scale in the experimental scans were the free
 parameters  in the analysis and varied relative to the calibrated energy 
 points. The beam energy spread varies in the range from 0.08 MeV at
 $\sqrt[]{s}= 440$ MeV to 0.35 MeV at $\sqrt[]{s}=970$ MeV.

\section{Data analysis}

\subsection{Selection of $e^+e^- \to \pi^+\pi^-\pi^0$ events}

 The data analysis and selection criteria used in this work are similar
 to those described in Ref.\cite{phi98,dplphi98,pi3mhad}. During the
 experimental runs, the first-level trigger \cite{sndnim} selects events with
 energy deposition in the calorimeter more than 180 MeV and with two or more
 charged particles. During processing of the experimental data the event
 reconstruction is performed \cite{sndnim,phi98}. For further analysis, events
 containing two or more photons and two charged particles with $|z| < 10$ cm
 and $r < 1$ cm were selected. Here $z$ is the coordinate of the charged
 particle production point along the beam axis (the longitudinal size of the
 interaction region depends on beam energy and varies from 2 to 2.5 cm); $r$
 is the distance between the charged particle track and the beam axis in the
 $r-\phi$ plane. Extra photons in $e^+e^- \to \pi^+\pi^-\pi^0$ events can
 appear because of the overlap with the beam background or nuclear
 interactions of the charged pions in the calorimeter. Under these selection
 conditions, the background sources are
 $e^+e^- \to \omega\pi^0 \to \pi^+\pi^-\pi^0\pi^0$,
 $e^+e^- \to e^+e^-\gamma$, $e^+e^-\gamma\gamma$, $\pi^+\pi^-(\gamma)$,
 $\mu^+\mu^-(\gamma)$ processes, cosmic and beam backgrounds.

 The polar angles of the charged particles were bounded by the criterion:
 $15^{\circ} < \theta < 165^{\circ}$.  To suppress the cosmic and beam
 backgrounds, the following cuts were applied: $E_{neu}>100$ MeV,
 $2\leq N_\gamma \leq 3$, $|\Delta z| < 3$ cm and $\psi > 20^{\circ}$.
 Here $E_{neu}$ is the energy deposition of the neutral particles, $N_\gamma$
 is the number of detected photons, $\Delta z = z_1-z_2,$ and $z_1,z_2$ are
 z-coordinates of the charged particles tracks, $\psi$ is the angle between two
 charged particles tracks.

 To suppress the  $e^+e^- \to e^+e^-\gamma\gamma$ events, an
 energy deposition of the charged particles in the calorimeter $E_{cha}$ was
 required to be small: $E_{cha} < 0.5 \cdot \sqrt[]{s}$.   

 To reject the background from the $e^+e^- \to \pi^+\pi^-(\gamma)$,
 $\mu^+\mu^-(\gamma)$ and $e^+e^-\gamma$, the following cut was imposed:
 $|\Delta\phi| > 5^\circ$. Here $\Delta\phi$ is an acollinearity angle of the
 charged particles in the azimuthal plane.

 For events left after these cuts, a kinematic fit was performed under the
 following constraints: the charged particles are assumed to be pions, the
 system has zero total momentum, the total energy is $\sqrt[]{s}$, and the
 photons originate from the $\pi^0 \to \gamma\gamma$ decays (Fig.\ref{mppi}).
 The value of the likelihood function $\chi^2_{3\pi}$ (Fig.\ref{xi2ua}) is
 calculated during the fit. In events with more than two photons, extra
 photons are considered as spurious ones and rejected. To do this, all
 possible subsets of two photons were inspected and the one, corresponding to
 the maximum likelihood was selected. The kinematic fits were also performed
 under assumptions that the $e^+e^-\to \pi^+\pi^-\gamma$,
 $e^+e^-\to\mu^+\mu^-\gamma$ or $e^+e^-\to e^+e^-\gamma$ events with extra
 photons were detected and the values of the corresponding likelihood functions
 $\chi^2_{2\pi\gamma}$, $\chi^2_{2\mu\gamma}$ and $\chi^2_{2e\gamma}$ were
 calculated. After the kinematic fits, the following cuts were applied:
 $\chi^2_{3\pi} < 20$, $\chi^2_{2\pi\gamma} > 20$, $\chi^2_{2\mu\gamma} > 20$
 and $\chi^2_{2e \gamma} > 20$, the polar angle $\theta_\gamma$ of at
 least one of the photons, selected by the reconstruction program as
 originated from the $\pi^0$ decay, should satisfy the following criterion:
 $36^{\circ} < \theta_\gamma < 144^{\circ}$.  In the energy
 region $\sqrt[]{s}<730$ MeV, for additional suppression of the background, the
 cut $p/E>0.5$ -- $0.3$ (at $\sqrt[]{s}=720$ -- $440$ MeV) was applied. Here
 $p$ and $E$ are the charged pions momentum and energy calculated after the
 kinematic reconstruction. For additional suppression of the
 $e^+e^-\to \omega\pi^0\to\pi^+\pi^-\pi^0\pi^0$ background, the criterion 
 $N_\gamma=2$ was applied for the energies $\sqrt[]{s}>900$ MeV.

 The angular distributions of particles for the selected events are shown in
 Fig.\ref{an12},\ref{teu12},\ref{teu3},\ref{sohe} and \ref{teu45}, while
 Fig.\ref{epu4},\ref{epu45},\ref{epu12} and \ref{epu3} demonstrate the photon
 and pion energy distributions for the same events.
 The experimental and simulated distributions are in agreement.

\subsection{Background subtraction}

 The number of background events was estimated from the following formula:
\begin{eqnarray}
\label{bg}
 N_{bkg}({s}) = \sum_i \sigma_{Ri}({s}) \epsilon_i({s})  IL({s}),
\end{eqnarray}
 where $i$ is a process number, $\sigma_{Ri}({s})$ is the cross
 section of the background process taking into account the radiative
 corrections, $IL({s})$ is the integrated luminosity, $\epsilon_i({s})$ is
 the detection probability for the background process obtained from simulation
 under selection criteria described above. For the 
 $e^+e^-\to\omega\pi$ background
 estimation, the cross section obtained in SND experiments \cite{ppg} was used.

 To estimate the accuracy of background events number determination, the
 $\chi^2_{3\pi}$ distribution in the energy region $\sqrt[]{s}>870$ MeV
 (Fig.\ref{xi2ufit}) was studied. The experimental $\chi^2_{3\pi}$
 distribution in the range $0<\chi^2_{3\pi}<40$ was fitted by a sum of
 background and signal. The distribution for background events was taken from
 the simulation, while for the signal $e^+e^-\to\pi^+\pi^-\pi^0$ events --
 by using experimental data collected in the vicinity of the 
 $\omega$ meson peak.
 As a result, the ratio between the number of background events obtained
 from the fit and the number calculated according to Eq.(\ref{bg}) was found
 to be $2.0 \pm 1.2$. The error was estimated by varying the selection
 criteria. Taking into account this ratio, the number of background events
 obtained from Eq.(\ref{bg}) was multiplied by a factor of 2 in all energy 
 points and the accuracy of the
 determination of the number of background events was estimated to be about
 60\%.

 The numbers of $e^+e^-\to \pi^+\pi^-\pi^0$ events (after the background
 subtraction) and background event numbers are shown in Table~\ref{tab1}. 
 
 To check the accuracy of background subtraction in the energy region
 $\sqrt[]{s}<730$ MeV, the data were analyzed in a different way. The 
 kinematic reconstruction was
 performed under the following constrains: the charged particles are pions, 
 the system has
 zero total momentum, the total energy is ~$\sqrt[]{s}$. The constrain that
 the photon originated from the $\pi^0\to\gamma\gamma$ decay was not used. In
 events with more than two photons, extra photons are considered as spurious
 ones and rejected. The value of the likelihood function 
 $\chi^2_{2\pi 2\gamma}$
 is calculated during the fit. After the fit, the following cut was applied:
 $\chi^2_{2\pi 2\gamma}<20$. For selected events the two-photon
 $m_{\gamma\gamma}$ invariant mass spectra (Fig.\ref{mpgg360}) were fitted by
 a sum of background and signal ($m_{\gamma\gamma}$ is calculated after the
 kinematic fit). The shape of the distribution for $e^+e^-\to\pi^+\pi^-\pi^0$
 events was obtained by using experimental data collected in the vicinity of 
 the $\omega$-resonance. For the background the uniform distribution was used
 (other assumptions about the shape of the background spectrum do not change 
 the fit results). 
 The cross sections obtained by using two different methods of background
 subtraction are in agreement (Table\ref{tab2}).
 The analogous check was performed for the energy region $\sqrt[]{s}>900$ MeV.
 The results of different approaches are again in agreement.
  
\subsection{Detection efficiency}

 The detection efficiency of the $e^+e^-\to\pi^+\pi^-\pi^0(\gamma)$
 process was obtained from simulation. To take into account the overlap of
 the beam background with the signal events, background events (experimental 
 events
 collected when the detector was triggered with an external generator) were
 mixed with the simulated events. The detection efficiency for events without
 $\gamma$-quantum radiation by the initial particles is about 0.35 (Table 
 \ref{tab1}). The detection efficiency dependence on the radiated photon energy
 is shown in Fig.\ref{efrad}. The efficiency decrease with the rise of the 
 radiated
 photon energy is due to the selection criterion $\chi^2_{3\pi}<20$, which
 involves the energy and momentum conservations in the
 $e^+e^-\to\pi^+\pi^-\pi^0$ process.

 Inaccuracies in the simulation of the distributions over some selection
 parameters lead to an error in the determination of average detection 
 efficiency.
 To take into account these uncertainties, the detection efficiency was
 multiplied by correction coefficients, which were obtained in the following
 way \cite{phi98,pi3mhad}. The experimental events were selected without any
 conditions on the parameter under study, using the selection parameters
 uncorrelated with the studied one. The same selection criteria were applied
 to simulated events. Then the cut was applied to the parameter and the
 correction coefficient was calculated: 
\begin{eqnarray}
 \delta = { {n/N} \over {m/M} },
\end{eqnarray}
 where $N$ and $M$ are the number of events in experiment and simulation
 respectively selected without any cuts on the parameter under study;
 $n$ and $m$ are the number of events in experiment and simulation when the
 cut on the parameter was applied. As a rule, the error in the coefficient
 $\delta$ determination is connected with the uncertainty of background
 subtraction. This systematic error was estimated by varying other selection
 criteria.

 The inaccuracy in $\chi^2_{3\pi}$ distribution simulation
 (Fig.\ref{xi2ua}) is the main source of uncertainty in the detection
 efficiency determination. The correction
 coefficient $\delta_{\chi^2_{3\pi}}=0.95\pm0.02$ (Fig.\ref{xi2ud}) was
 obtained by using data collected in the vicinity of the $\omega$ resonance. 
 The
 error due to uncertainty in simulation of other parameters is estimated to be
 1.7\%. In the energy region $\sqrt[]{s}>900$ MeV the additional selection
 criterion $N_\gamma=2$ was applied. The uncertainty due to this cut was
 estimated to be 3\%. The systematic error of the detection efficiency
 determination is 2.7\% in the energy region $\sqrt[]{s}<900$ MeV and is about
 4.1\% at $\sqrt[]{s}>900$. The detection efficiency after the applied 
 corrections is shown in Table~\ref{tab1}.

\section{Theoretical framework}

 In the framework of the vector meson dominance model, the cross section
 of the  $e^+e^-\to\pi^+\pi^-\pi^0$ process is
\begin{eqnarray}
\label{ds}
 {d\sigma \over dm_0 dm_+} = { {4\pi\alpha} \over {s^{3/2}} }
 {{|\vec{p}_+ \times \vec{p}_-|^2} \over {12\pi^2\mbox{~}\sqrt[]{s}}} m_0m_
 + \cdot |F|^2,
\end{eqnarray}
 where $\vec{p}_+$ and $\vec{p}_-$ are the $\pi^+$ and $\pi^-$ momenta,
 $m_0$ and $m_+$ are  $\pi^+\pi^-$ and $\pi^+\pi^0$ pairs invariant masses. 
 The form factor $F$ of the $\gamma^\star \to \pi^+\pi^-\pi^0$ transition
  has the form
\begin{eqnarray}
\label{formfac}
 |F|^2 = \Biggl| A_{\rho\pi}(s)  \sum_{i=+,0,-} 
 { g_{\rho^i\pi\pi} \over {D_\rho(m_i)Z(m_i)}} +
 A_{\rho^{\prime(\prime\prime)}\pi}(s)  \sum_{i=+,0,-}
 { g_{\rho^{\prime(\prime\prime)}\pi\pi} \over 
 {D_{\rho^{\prime(\prime\prime)}}(m_i)}}  + 
 A_{\omega\pi}(s)
 {\Pi_{\rho\omega}g_{\rho^0\pi\pi}\over D_\rho(m_0) D_\omega(m_0)} \Biggr|^2
\end{eqnarray}
 The first term in Eq.(\ref{formfac}) takes into account the  
 $\gamma^\star\to \rho\pi\to\pi^+\pi^-\pi^0$ transition
 (Fig.\ref{dplfeiman} a), which dominates in the process under study
 \cite{dplphi98}.
 Here
 $$D_\rho(m_i) = m_{\rho^i}^2 - m_i^2 -im_i\Gamma_{\rho^i}(m_i), \mbox{~~~}
 \Gamma_{\rho^i}(m_i) = \Biggl({m_{\rho^i} \over m_i}\Biggr)^2
   \Gamma_{\rho^i} \cdot \Biggl({q_i(m_i) \over q_i(m_{\rho^i})}\Biggr)^3$$
 $$q_0(m^2) = {1 \over 2}(m^2-4m_\pi^2)^{1/2}, \mbox{~~~}
 q_\pm(m^2) = {1 \over 2m}
 \bigl[(m^2-(m_{\pi^0}+m_\pi)^2)(m^2-(m_{\pi^0}-m_\pi)^2)\bigr]^{1/2},$$
 $$m_-=\sqrt[]{s+m_{\pi^0}^2+2m_{\pi}^2-m_0^2-m_+^2},$$
 where $m_-$ is the $\pi^-\pi^0$ pair invariant mass, $m_{\pi^0}$ and 
 $m_\pi$ are
 the neutral and charged pion masses, $i$ denotes the sign of the $\rho$-meson
 ($\pi\pi$ pair) charge. The $\rho^0 \to \pi^+\pi^-$ and
 $\rho^\pm \to \pi^\pm\pi^0$ transition coupling constants could be determined
 in the following way:
 $$g_{\rho^0\pi\pi}^2 = {6\pi m_{\rho^0}^2\Gamma_{\rho^0} \over
 q_0(m_{\rho^0})^3}, \mbox{~~}
 g_{\rho^\pm\pi\pi}^2 = {6\pi m_{\rho^\pm}^2\Gamma_{\rho^\pm} \over
 q_\pm(m_{\rho^\pm})^3}$$
 Experimental data \cite{dplphi98} do not contradict to the equality of the
 coupling constants
 $g_{\rho^0\pi\pi}^2 = g_{\rho^\pm\pi\pi}^2$. In this case the $\rho^0$ and
 $\rho^\pm$ meson widths are related as follows:
\begin{eqnarray}
\label{shir}
 \Gamma_{\rho^\pm} = \Gamma_{\rho^0}{m_{\rho^0}^2 \over m_{\rho^\pm}^2}
 { q_\pm(m_{\rho^\pm})^3 \over q_0(m_{\rho^0})^3}.
\end{eqnarray}
 In the subsequent analysis we assume that
 $g_{\rho^0\pi\pi}^2 = g_{\rho^\pm\pi\pi}^2$, and the width values were
 taken from the SND measurements \cite{dplphi98}: 
 $\Gamma_{\rho^0} = 149.8$ MeV,
 $\Gamma_{\rho^\pm} = 150.9$ MeV. The neutral and charged $\rho$ mesons masses
 were assumed to be equal and were also taken from the SND measurements
 \cite{dplphi98} $m_\rho=775.0$ MeV.
 A factor $Z(m) = 1 -is_1\Phi(m,s)$ takes into account the interaction of the
 $\rho$ and $\pi$ mesons in the final state \cite{akfaz} 
 (Fig.\ref{dplfeiman} d),
 parameter $s_1=1\pm 0.2$ corresponds to the prediction of Ref.\cite{akfaz},
 where the concrete form of the $\Phi(m,s)$ function can be found.
 In experimental studies of the $\pi\pi$ mass spectra in the $e^+e^-\to 3\pi$
 process at $\sqrt[]{s}\simeq m_\phi$ \cite{dplphi98} we had obtained
 $s_1=0.3\pm0.3\pm0.3$. This result is consistent with zero, but also does not
 contradict to the prediction of Ref. \cite{akfaz}.

 The second term in Eq.(\ref{formfac}) takes into account the possible
 transition $\gamma^\star\to\rho^{\prime(\prime\prime)}\pi\to\pi^+\pi^-\pi^0$
 (Fig.\ref{dplfeiman} c). This term can be written as
 $A_{\rho\pi}(s) \cdot a_{3\pi}$, where
$$
 a_{3\pi} =  {A_{\rho^{\prime(\prime\prime)}\pi}(s) \over A_{\rho\pi}(s)} 
 \sum_{i=+,0,-} { g_{\rho^{\prime(\prime\prime)}\pi\pi} \over
 {D_{\rho^{\prime(\prime\prime)}}(m_i)}}.
$$
 In the analysis, the $a_{3\pi}$ amplitude was assumed to be a real constant.
 From the
 $\pi\pi$ mass spectra analysis, in the process $e^+e^-\to 3\pi$ at 
 $\sqrt[]{s}\simeq m_\phi$ \cite{dplphi98}, it was found that
 $a_{3\pi} = (0.01 \pm 0.23 \pm 0.25) \times 10^{-5}$ MeV$^{-2}$.

 The third term in Eq.(\ref{formfac}) takes into account the
 $\gamma^\star\to\omega\pi^0\to\pi^+\pi^-\pi^0$ transition \cite{thrhoom}
 (Fig.\ref{dplfeiman} b). The polarization operator of the $\rho-\omega$ mixing
 satisfies $\mbox{Im}(\Pi_{\rho\omega}) \ll \mbox{Re}(\Pi_{\rho\omega})$
 \cite{akfaz,akozi}, where
\begin{eqnarray}
 \mbox{Re}(\Pi_{\rho\omega}) =
 \sqrt[]{{\Gamma_\omega \over \Gamma_{\rho^0}(m_\omega)} 
 B(\omega\to\pi^+\pi^-)} \cdot \biggl| (m_\omega^2-m_\rho^2) - 
 im_\omega(\Gamma_\omega - \Gamma_{\rho^0}(m_\omega))\biggr|,
\end{eqnarray}
 so we have assumed $\mbox{Im}(\Pi_{\rho\omega}) = 0$ in the subsequent 
 analysis.

 The $e^+e^-\to\pi^+\pi^-\pi^0$ process cross section can be written in the
 following form:
\begin{eqnarray}
\label{sech3p}
 \sigma_{3\pi} = \sigma_{\rho\pi\to3\pi} + \sigma_{\omega\pi\to3\pi} +
 \sigma_{int},
\end{eqnarray}
 where
\begin{eqnarray}
\label{sech1}
 \sigma_{\rho\pi\to3\pi} = {{4\pi\alpha} \over {s^{3/2}}}
 W_{\rho\pi}(s)\biggl| A_{\rho\pi}(s) \biggr|^2,
\end{eqnarray}
\begin{eqnarray}
\label{sech2}
 \sigma_{\omega\pi\to3\pi} = {{4\pi\alpha} \over {s^{3/2}}}
 W_{\omega\pi}(s)\biggl| A_{\omega\pi}(s) \biggr|^2,
\end{eqnarray}
\begin{eqnarray}
\label{sech3}
 \sigma_{int} = {{4\pi\alpha} \over {s^{3/2}}}
 \biggl\{ A_{\rho\pi}(s) A_{\omega\pi}^\star (s) W_{int}(s) +
 A_{\rho\pi}^\star (s) A_{\omega\pi}(s) W_{int}^\star (s) \biggr\}.
\end{eqnarray}
 The phase space factors  $W_{\rho\pi}(s)$, $W_{\omega\pi}(s)$ and $W_{int}(s)$
 were calculated as follows:
\begin{eqnarray}
 W_{\rho\pi}(s) = {1 \over 12 \pi^2 \mbox{~}\sqrt[]{s}}
 \int\limits^{\sqrt[]{s}-m_{\pi^0}}_{2m_\pi}
 m_0 dm_0 \int\limits^{m_+^{max}(m_0)}_{m_+^{min}(m_0)}
 m_+ dm_+ |\vec{p}_+ \times \vec{p}_-|^2 \cdot \biggl|\sum_{i=+,0,-} 
 { g_{\rho^i\pi\pi} \over D_\rho(m_i) Z(m_i)}+a_{3\pi}\biggr|^2,
\end{eqnarray}
\begin{eqnarray}
 W_{\omega\pi}(s) = {1 \over 12 \pi^2 \mbox{~}\sqrt[]{s}}
 \int\limits^{\sqrt[]{s}-m_{\pi^0}}_{2m_\pi}
 m_0 dm_0 \int\limits^{m_+^{max}(m_0)}_{m_+^{min}(m_0)}
 m_+ dm_+ |\vec{p}_+ \times \vec{p}_-|^2 \cdot
 \biggl| {\Pi_{\rho\omega}g_{\rho^0\pi\pi}\over D_\rho(m_0) D_\omega(m_0)}
 \biggr|^2,
\end{eqnarray}
\begin{eqnarray}
\nonumber
{ W_{int}(s) = {1 \over 12 \pi^2 \mbox{~}\sqrt[]{s}}
 \int\limits^{\sqrt[]{s}-m_{\pi^0}}_{2m_\pi}
 m_0 dm_0 \int\limits^{m_+^{max}(m_0)}_{m_+^{min}(m_0)}
 m_+ dm_+ |\vec{p}_+ \times \vec{p}_-|^2 
 \biggl[ {\Pi_{\rho\omega}g_{\rho^0\pi\pi}\over 
 D_\rho(m_0) D_\omega(m_0)}\biggr]^\star \times} \\
 \times \biggl[ \sum_{i=+,0,-}{ g_{\rho^i\pi\pi}\over D_\rho(m_i) Z(m_i)} +
 a_{3\pi}
 \biggr]
\end{eqnarray}
  
 Amplitudes of the $\gamma^\star\to\rho\pi$ transition have the form
\begin{eqnarray}
\label{aropi}
 A_{\rho\pi}(s) = {{1}\over\sqrt[]{4\pi\alpha}}
 \sum_{V=\omega,\rho,\phi,\omega^\prime,\omega^{\prime\prime}}
 {{\Gamma_V m_V^2 \mbox{~} \sqrt[]{m_V\sigma(V\to 3\pi)}}\over{D_V(s)}}
 {{e^{i\phi_{\omega V}}C_{V\rho\pi}(s,r_0)}\over{\sqrt[]{W_{\rho\pi}(m_V)}}},
\end{eqnarray}
 where
$$D_V(s)=m_V^2-s-i\mbox{~}\sqrt[]{s}\Gamma_V(s), \mbox{~~~}
  \Gamma_V(s)=\sum_{f}\Gamma(V\to f,s).$$
 Here $f$ denotes the final state of the vector meson $V$ decay, $m_V$ is the
 vector meson mass, $\Gamma_V=\Gamma_V(m_V)$. The $\phi$ meson mass and width
 were taken from the SND measurements $m_\phi=1019.42$ MeV, $\Gamma_\phi=4.21$
 MeV \cite{phi98}. The following forms of the energy dependence of the vector 
 mesons total widths were used
$$ \Gamma_\rho(s)={m_\rho^2 \over s}{q_0(s)^3 \over q_0(m_\rho)^3}\Gamma_\rho
   C^2_{\rho\pi\pi}(s,r_0) +
 {{g_{\rho\omega\pi}^2}\over{12\pi}}q^3_{\omega\pi}(s),$$
$$ \Gamma_\omega(s) = {m_\omega^2 \over s}{q_0(s)^3 \over q_0(m_\omega)^3}
   \Gamma_\omega B(\omega\to\pi^+\pi^-) C^2_{\omega \pi\pi}(s,r_0) +
   {q_{\pi\gamma}(s)^3 \over q_{\pi\gamma}(m_\omega)^3}
   \Gamma_\omega B(\omega\to\pi^0\gamma) C^2_{\omega\gamma\pi}(s,r_0) +$$
$$ +   {W_{\rho\pi}(s) \over W_{\rho\pi}(m_\omega)}
   \Gamma_\omega B(\omega\to 3\pi) C^2_{\omega\rho\pi}(s,r_0),$$
$$ \Gamma_\phi(s) = {m_\phi^2 \over s}
   {q_{K^\pm}(s)^3 \over q_{K^\pm}(m_\phi)^3}
   \Gamma_\phi B(\phi\to K^+K^-) C^2_{\phi K^+K^-}(s,r_0) +
   {m_\phi^2 \over s}
   {q_{K^0}(s)^3 \over q_{K^0}(m_\phi)^3}
   \Gamma_\phi B(\phi\to K_SK_L) C^2_{\phi K_SK_L}(s,r_0) + $$
$$ +   {q_{\eta\gamma}(s)^3 \over q_{\eta\gamma}(m_\phi)^3}
   \Gamma_\phi B(\phi\to\eta\gamma) C^2_{\phi\gamma\eta}(s,r_0) +
   {W_{\rho\pi}(s)\over W_{\rho\pi}(m_\phi)}
   \Gamma_\phi B(\phi\to 3\pi) C^2_{\phi\rho\pi}(s,r_0),$$
$$ \Gamma_{\omega^\prime}(s) = 
   \Gamma_{\omega^\prime} C^2_{\omega^\prime\rho\pi}(s,r_0)
 {W_{\rho\pi}(s) \over W_{\rho\pi}(m_{\omega^{\prime}})},$$ 
$$\Gamma_{\omega^{\prime\prime}}(s) =
  \Gamma_{\omega^{\prime\prime}}C^2_{\omega^{\prime\prime}\rho\pi}(s,r_0)
  \biggl(
  {W_{\rho\pi}(s) \over W_{\rho\pi}(m_{\omega^{\prime\prime}})}
  B(\omega^{\prime\prime}\to3\pi) +
  {W_{\omega\pi\pi}(s) \over W_{\omega\pi\pi}(m_{\omega^{\prime\prime}})}
  B(\omega^{\prime\prime} \to \omega\pi\pi)\biggr).$$
 Here $g_{\rho\omega\pi}$ is a coupling constant of $\rho\to\omega\pi^0$
 transition, $q_{\omega\pi}$, $q_{K^\pm}$, $q_{K^0}$, $q_{\pi\gamma}$ and
 $q_{\eta\gamma}$ are the $\omega$ meson, kaon, $\eta$ meson and pion momenta,
 $W_{\omega\pi\pi}(s)$ is the phase space factor of the $\omega\pi\pi$ final
 state \cite{ak2}, $C_{VPP}(s,r_0)$ and $C_{VVP}(s,r_0)$ are the form factors
 which restrict too fast growth with energy of the partial widths, so that
 $\sqrt[]{s}\Gamma(s)\to const$ as $s\to \infty$.
 According to Ref.\cite{freshlook} these form factors can be written as follows
\begin{eqnarray}
\label{cform}
C_{V\gamma(\rho)P}(s,r_0)={{1+(r_0m_V)^2} \over {1+(r_0\sqrt[]{s})^2}},
   \mbox{~~~}
   C_{VPP}(s,r_0)=\sqrt[]{{1+(r_0q_P(m_V))^2} \over {1+(r_0q_P(s))^2}},
\end{eqnarray}
 where $q_P$ is the momentum of the pseudoscalar meson, $r_0$ is the range 
 parameter (its value was taken to be the same for all decays).

 The relative probabilities  of the decays were calculated as follows 
 $$B(V\to X)={\sigma(V\to X)\over\sigma(V)}, \mbox{~~}
 \sigma(V)=\sum_{X} \sigma(V\to X),  \mbox{~~}
 \sigma(V\to X) = {{12\pi B(V\to e^+e^-)B(V\to X) } \over {m_V^2}}.$$
 In particular:
 $$B(\omega\to X)={\sigma(\omega\to X) \over \sigma(\omega)},
 \mbox{~~} \sigma(\omega)=
 {{\sigma(\omega\to3\pi)+\sigma(\omega\to\pi^0\gamma)} \over
 {1-B(\omega\to\pi^+\pi^-)}}.$$
 In further analysis we have used $\sigma(\omega\to\pi^0\gamma)=155.8$ nb,
 $\sigma(\phi\to K^+K^-)=1968$ nb, $\sigma(\phi\to K_SK_L)=1451$ nb and
 $\sigma(\phi\to\eta\gamma)=54.8$ nb obtained in the SND experiments
 \cite{pi0gam,phi98,etagam}.

 $\phi_{\omega V}$ is a relative interference phase between the vector meson 
 $V$ and
 $\omega$, so $\phi_{\omega\omega}=0^\circ$. Phases $\phi_{\omega V}$ can
 deviate from $-180^\circ$ or $180^\circ$ and their values can be energy
 dependent due to mixing between vector mesons. For example, the phase
 $\phi_{\omega\phi}$ was found to be close to $180^\circ$ \cite{phi98,pi3mhad}
 in agreement with the prediction \cite{faza} $\phi_{\omega\phi}=\Phi(s)$
 $(\Phi(m_\phi)\simeq 163^\circ)$, where the function $\Phi(s)$ is defined in
 Ref.\cite{faza}. In Ref.\cite{pi3mhad} it was shown that
 $\phi_{\omega\omega^\prime} \sim 180^\circ$ and
 $\phi_{\omega\omega^{\prime\prime}} \sim 0^\circ$, so in this work these two
 phases were fixed on those values.

 Taking into account the $\rho-\omega$ mixing, the $\omega\to\rho\pi$ and
 $\rho\to\rho\pi$ transition amplitudes can be written in the following way
 \cite{thrhoom,akozi}
\begin{eqnarray}
 \label{rhoom}
 A_{\omega\to\rho\pi} + A_{\rho\to\rho\pi} =
 { {g^{(0)}_{\gamma\omega}g^{(0)}_{\omega\rho\pi}} \over {D_{\omega}(s)} }
 \biggl[1+{g^{(0)}_{\gamma\rho}\over g^{(0)}_{\gamma\omega}}\varepsilon(s) 
 \biggr]+
 {{g^{(0)}_{\gamma\rho}g^{(0)}_{\omega\rho\pi}}\over{D_{\rho}(s)}}
 \biggl[{g^{(0)}_{\rho\rho\pi}\over g^{(0)}_{\omega\rho\pi}}-\varepsilon(s)
 \biggr],
\end{eqnarray}
 where
 $$\varepsilon(s) = {-\Pi_{\rho\omega} \over {D_\omega(s)-D_\rho(s)}}, 
  \mbox{~~}
 |g_{V\gamma}| = \Biggl[ {{3m_V^3\Gamma_VB(V \to e^+e^-)} \over
 {4\pi\alpha}} \Biggr]^{1/2}, \mbox{~~}
 |g_{V\rho\pi}| = \Biggl[{{4\pi\Gamma_VB(V \to \rho\pi)} \over
 {W_{\rho\pi}(m_V)}} \Biggr]^{1/2}.$$
 The superscript $(0)$ denotes the coupling constants of the pure, unmixed
 state. The Eq.(\ref{rhoom}) can be rewritten as follows
$$A_{\omega\to\rho\pi} + A_{\rho\to\rho\pi} = {{1}\over\sqrt[]{4\pi\alpha}}   
 \sum_{V=\omega,\rho}
  { {\Gamma_V m_V^2 \mbox{~} \sqrt[]{m_V\sigma(V\to 3\pi)} } \over {D_V(s)} } 
  { f_{V\rho\pi}(s)C_{V\rho\pi}(s,r_0) \over {\sqrt[]{W_{\rho\pi}(m_V)}} },
  $$
 where
 $$f_{V\rho\pi}(s) = {r_{V\rho\pi}(s) \over r_{V\rho\pi}(m_V) }, \mbox{~~}$$
 $$r_{\omega\rho\pi}(s) =
 1+{g^{(0)}_{\gamma\rho}\over g^{(0)}_{\gamma\omega}}\varepsilon(s) \simeq
 1+\biggl[{m_\rho^3 \Gamma(\rho\to e^+e^-) \over
 m_\omega^3 \Gamma(\omega\to e^+e^-)}\biggr]^{1/2} \varepsilon(s), \mbox{~~}$$
$$ r_{\rho\rho\pi}(s) =
 {g^{(0)}_{\rho\rho\pi}\over g^{(0)}_{\omega\rho\pi}}-\varepsilon(s) \simeq
 -\varepsilon(s)$$
 If the $\rho\to 3\pi$ transition proceeds only via $\rho-\omega$ mixing, that
 is $g^{(0)}_{\rho\rho\pi}=0$, then
 $\phi_{\omega\rho} \simeq -90^\circ$ and almost does not depend on energy,
 besides
 $$\sigma(\rho\to 3\pi) \simeq \sigma(\omega\to 3\pi) 
 {m_\omega^2 \over m_\rho^2}{\Gamma_\omega \over \Gamma_\rho}
 {W_{\rho\pi}(m_\rho) \over W_{\rho\pi}(m_\omega)} 
 {r_{\rho\rho\pi}(m_\rho) \over r_{\omega\rho\pi}(m_\omega)}
 {B(\rho\to e^+e^-)\over B(\omega\to e^+e^-)}.$$

 For the $\gamma^\star\to\omega\pi^0$ transition amplitude, the model which 
 gives satisfactory description of the relative phase between it and the 
 $A_{\rho\pi}(s)$,
 Eq.(\ref{aropi}), amplitude \cite{pi3mhad} was used:
\begin{eqnarray}
\label{aompi}
 A_{\omega\pi}(s) =\sqrt[]{ { 3 \over 4\pi\alpha }} \times \Biggl[  \mbox{~}
 {\sqrt[]{m_\rho^3 \Gamma_\rho B(\rho\to e^+e^-)} g_{\rho\omega\pi} \over
  D_\rho(s)} +
  \sum_{V=\rho^\prime,\rho^{\prime\prime}} 
   { {\Gamma_V m_V^2 \mbox{~} \sqrt[]{m_V\sigma(V\to \omega\pi^0)}} \over 
   D_V(s) }  
   { e^{i\phi_{\rho V}} \over \sqrt[]{q_{\omega\pi}^3(m_V)}} \mbox{~} \Biggr],
\end{eqnarray}
 where $\phi_{\rho\rho}=0^\circ$, $m_{\rho^\prime}=1480$ MeV,
 $\Gamma_{\rho^\prime}=790$ MeV, $\sigma(\rho^\prime\to\omega\pi^0)=86$ nb,
 $\phi_{\rho^\prime}=180^\circ$, $m_{\rho^{\prime\prime}}=1640$ MeV,
 $\Gamma_{\rho^{\prime\prime}}=1290$ MeV,
 $\sigma(\rho^{\prime\prime}\to\omega\pi^0)=48$ nb, 
 $\phi_{\rho^\prime}=0^\circ$,
 $g_{\rho\omega\pi}=16,8$ GeV$^{-1}$ and
 $$\Gamma_{\rho^{\prime(\prime\prime)}}(s) = 
   \Gamma_{\rho^{\prime(\prime\prime)}} 
   {q_{\omega\pi}^3(s)\over q_{\omega\pi}^3(m_V)}.$$

\section{Cross section measurement}

 From the data in Table~\ref{tab1}, the cross section of the process
 $e^+e^-\to\pi^+\pi^-\pi^0$ can be calculated as follows:
\begin{eqnarray}
\label{aprox}
 \sigma(s) = {{N_{3\pi}(s)} \over {IL(s)\xi(s)}},
\end{eqnarray}
 where $N_{3\pi}(s)$ is the number of selected
 $e^+e^-\to\pi^+\pi^-\pi^0(\gamma)$ events, $IL(s)$ is the integrated
 luminosity, $\xi(s)$ is the function which takes into account the detection
 efficiency and radiative corrections for the initial state radiation:
\begin{eqnarray}
\label{xifu}
 \xi(s) = {\int\limits^{E^{max}_\gamma}_0 \sigma_{3\pi}(s,E_\gamma)
 F(s,E_\gamma)
 \epsilon(s,E_\gamma) \mathrm{d}E_\gamma \over {\sigma_{3\pi}(s)}}.
\end{eqnarray}
 Here $E_\gamma$ is the emitted photon energy, $F(s,E_\gamma)$ is the electron
 ``radiator'' function \cite{fadin}, $\epsilon(s,E_\gamma)$ is the detection
 efficiency of the process $e^+e^-\to\pi^+\pi^-\pi^0(\gamma_{rad})$ as a
 function of the emitted photon energy and the total energy in the $e^+e^-$
 center of mass system, $\sigma_{3\pi}(s)$ is the theoretical energy dependence
 of the cross section given by the equation (\ref{sech3p}).

 To obtain the values of $\xi(s)$ at each energy point, the visible cross
 section of the process $e^+e^-\to\pi^+\pi^-\pi^0(\gamma_{rad})$
 $$\sigma^{vis}(s) = {N_{3\pi}(s) \over IL(s)}$$
 was fitted by theoretical energy dependence
 $$\sigma^{th}(s) = \sigma_{3\pi}(s)\xi(s).$$
 The following logarithmic likelihood function was minimized:
 $$\chi^2=\sum_{i} {{(\sigma^{vis}_i-\sigma^{th}_i)^2}\over{\Delta^2_i}},$$
 where $i$ is the energy point number, $\Delta_i$ is the error of the visible
 cross section $\sigma^{vis}$.

 In the fit the $\phi$, $\omega^\prime$, $\omega^{\prime\prime}$ meson
 parameters(the mass, width, branching ratios of main decays ) was fixed at 
 their values obtained in the SND experiments \cite{phi98,pi3mhad}, other 
 parameters were fixed as follows $r_0=0$,
 $a_{3\pi}=0$, $s_1=0$ and $\phi_{\omega\phi}=\Phi(s)$
 $(\Phi(m_\phi)\simeq 163^\circ)$. The Eq.(\ref{aropi}) was written in the
 following form:
\begin{eqnarray}
 A_{\rho\pi}(s) = {{1}\over\sqrt[]{4\pi\alpha}}
 \sum_{V=\omega,\rho,\phi,\omega^\prime,\omega^{\prime\prime}}
 {{\Gamma_V m_V^2 \mbox{~} \sqrt[]{m_V\sigma(V\to 3\pi)}}\over{D_V(s)}}
 {{e^{i\phi_{\omega V}}}\over{\sqrt[]{W_{\rho\pi}(m_V)}}} + C_{3\pi},
\end{eqnarray}
 where $C_{3\pi}$ is a complex constant.

 The $\sigma(\omega\to 3\pi)$, $\Gamma_{\omega}$ and $m_{\omega}$ were the
 free parameters of the fit. The $\xi(s)$  values were obtained from the
 approximation of the experimental data  in several models:
\begin{enumerate}
\item
 $\sigma(\rho\to 3\pi)=0$, $C_{3\pi}=0$
\item
 $\sigma(\rho\to 3\pi)$ and $\phi_{\omega\rho}$ are free parameters,
 $C_{3\pi}=0$
\item
 $C_{3\pi}$ is a free parameter, $\sigma(\rho\to 3\pi)=0$
\item
 $\sigma(\rho\to 3\pi)$, $\phi_{\omega\rho}$ and $C_{3\pi}$ are free parameters
\end{enumerate}
  The fits were also performed under the same assumptions, but with
  $\sigma(\omega^\prime\to 3\pi)=0$ and $\sigma(\omega^{\prime\prime}
  \to 3\pi)=0$.

 The values of $\xi(s)$ actually do not depend on the applied model.
 The largest model dependence, about $1.5-2 \%$, was found at $\sqrt[]{s}$
 from 800 to 840 MeV. Using the obtained $\xi(s)$ values, the cross section 
 of the $e^+e^-\to\pi^+\pi^-\pi^0$ process was calculated (Table~\ref{tab3} ).
 The systematic error of the cross section determination at each energy point
 $\sqrt[]{s}$ is equal to
 $$ 
 \sigma_{sys} = \sigma_{eff} \oplus \sigma_{IL} \oplus \sigma_{mod}(s) \oplus
 \sigma_{bkg}(s).
 $$
 Here $\sigma_{eff}=2.7\%$ at $\sqrt[]{s}<900$ MeV and
 $\sigma_{eff}=4.1\%$ at $\sqrt[]{s}>900$ MeV, $\sigma_{IL}=2\%$. They are
 systematic uncertainties in the detection efficiency and integrated
 luminosity, which are common for all energy points. The model uncertainty
 $\sigma_{mod}(s)$ was obtained from the difference of $\xi(s)$ values
 determined for the models mentioned above. The error $\sigma_{bkg}(s)$ takes
 into account the inaccuracy ($\sim 60\%$) of the background subtraction and
 depends on the beam energy.

\section{The \lowercase{\boldmath{$e^+e^- \to\pi^+\pi^-\pi^0$}} cross section
         analysis.}

 The cross section measured in this work (Table~\ref{tab3}) was analyzed
 together with the $e^+e^-\to\pi^+\pi^-\pi^0$ cross section measured by SND
 in the energy region $\sqrt[]{s}$ from 980 up to 1380 MeV \cite{phi98,pi3mhad}
 and with the DM2 results of the $e^+e^-\to\pi^+\pi^-\pi^0$ and
 $e^+e^-\to\omega\pi^+\pi^-$ cross sections measurements in the energy region
 $\sqrt[]{s}$ from 1340 to 2200 MeV \cite{dm2}.

 The $e^+e^-\to \pi^+\pi^-\pi^0$ cross section was fitted by the expression 
 (\ref{sech3p}). The $e^+e^-\to\omega\pi^+\pi^-$ process cross section was
 written in the following way:
\begin{eqnarray}
 \sigma_{\omega\pi\pi} = {1 \over s^{3/2}} \Biggl|
 {{\Gamma_{\omega^{\prime\prime}}m_{\omega^{\prime\prime}}^2
 \mbox{~}\sqrt[]{\sigma(\omega^{\prime\prime}\to\omega\pi^+\pi^-)
 m_{\omega^{\prime\prime}}}} \over{D_{\omega^{\prime\prime}(s)}}} \mbox{~}
 \sqrt[]{{W_{\omega\pi\pi}(s) C_{VVP}(s,r_0)}\over
 {W_{\omega\pi\pi}(m_{\omega^{\prime\prime}})}} \Biggr|^2.
\end{eqnarray}
 Here we have neglected the $\omega^{\prime}\to \omega\pi\pi$ contribution.
 To calculate $B(\omega^{\prime\prime}\to\omega\pi\pi)$, we have assumed
 $\sigma(\omega^{\prime\prime}\to\omega\pi\pi) = 1.5 \cdot \sigma(\omega^{\prime\prime}\to\omega\pi^+\pi^-)$.

 The cross sections of the $e^+e^-\to \pi^+\pi^-\pi^0$ and $\omega\pi^+\pi^-$
 processes, measured by SND and DM2, were fitted together.
 The function to be minimized was
 $$\chi^2_{tot}=\chi^2_{3\pi(SND)}+\chi^2_{3\pi(DM2)}+ \chi^2_{\omega\pi\pi(DM2)},$$
 where
 $$\chi^2_{3\pi(SND)} = \sum_{s}\Biggl( 
 { {\sigma_{3\pi}^{(SND)}(s)-\sigma_{3\pi}(s)} \over
 {\Delta_{3\pi}^{(SND)}(s)} } \Biggr)^2$$
 $$\chi^2_{3\pi(DM2)} = \sum_{s}\Biggl( 
 { C\cdot{\sigma_{3\pi}^{(DM2)}(s)-\sigma_{3\pi}(s)} \over
 {\Delta_{3\pi}^{(DM2)}(s)} }
 \Biggr)^2$$
 $$\chi^2_{\omega\pi\pi(DM2)} = \sum_{s}\Biggl(
 { C\cdot{\sigma_{\omega\pi\pi}^{(DM2)}(s)-
 \sigma_{\omega\pi\pi}(s)} \over {\Delta_{\omega\pi\pi}^{(DM2)}(s)} }
 \Biggr)^2$$
 Here $\sigma_{3\pi(\omega\pi\pi)}^{[SND(DM2)]}(s)$ are the experimental
 cross sections, $\Delta$ are their uncertainties, $C$ is a coefficient which
 take into account the relative systematic bias between SND and DM2 data.
 The errors $\Delta_{3\pi}^{(SND)}$ include both the statistical 
 $\sigma_{stat}$ and the systematic errors:
 $\Delta_{3\pi}^{(SND)}=\sigma_{stat}\oplus\sigma_{mod}\oplus\sigma_{bkg}$.
 In Ref.\cite{pi3mhad} the $C$ coefficient was estimated to be $C=1.54$.  
 In the analysis that follows we have fixed this coefficient at 1 or 1.54.

 In the fittings $m_\omega$, $\Gamma_\omega$, $\sigma(\omega\to 3\pi)$,
 $\sigma(\phi\to 3\pi)$, $\phi_{\omega\phi}$,
 $m_{\omega^\prime}$, $\Gamma_{\omega^\prime}$,
 $\sigma(\omega^\prime\to 3\pi)$,
 $m_{\omega^{\prime\prime}}$, $\Gamma_{\omega^{\prime\prime}}$,
 $\sigma(\omega^{\prime\prime}\to 3\pi)$ and
 $\sigma(\omega^{\prime\prime}\to\omega\pi^+\pi^-)$ were the free parameters.
 The approximations were performed under the following assumptions
 about the phase space factor for the $\pi^+\pi^-\pi^0$ final state:
\begin{enumerate}                                                              
\item
 $s_1=0$, $a_{3\pi}=0$;
\item
 $s_1=1$, $a_{3\pi}=0$;
\item
 $s_1=0$, $a_{3\pi}=-4\times 10^{-6}$ MeV$^{-2}$;
\item
 $s_1=0$, $a_{3\pi}= 4\times 10^{-6}$ MeV$^{-2}$.
\end{enumerate}
 The nonzero values of the $a_{3\pi}$ amplitude are the upper limits imposed 
 on the 90\% confidence level by using the SND result reported in 
 Ref.\cite{dplphi98}.
 The approximations were also performed without taking into account the 
 contribution from the $e^+e^-\to\omega\pi^0\to\pi^+\pi^-\pi^0$ process, i.e.
 by assuming $\sigma_{3\pi}=\sigma_{\rho\pi}$. The difference in the fit 
 results was included in the model uncertainty.
 
 The fittings were performed under the following model parameters: 
\begin{enumerate}
\item
 $r_0=0$, $\sigma(\rho\to 3\pi)=0$;
\item
 $r_0=0$, $\phi_{\omega\rho}$ and $\sigma(\rho\to 3\pi)$ are free parameters;
\item
 $\sigma(\rho\to 3\pi)=0$, $r_0$ is a free parameter.
\end{enumerate}
 The results of the fits are shown in Table~\ref{tab4} and in Fig.\ref{crf1},
 \ref{crf2},\ref{crf3},\ref{crf4},\ref{crf5},\ref{crf6}.
 In the Table~\ref{tab4}, $\chi^2_\omega$, $\chi^2_{(SND)}$
 ($\chi^2_{3\pi(SND)}=\chi^2_\omega+\chi^2_{(SND)}$) and $\chi^2_{(1)}$
 denote the $\chi^2$ values for the energy regions $\sqrt[]{s}$ below and
 above 970 MeV and in the energy range $880 \leq \sqrt[]{s} \leq 970$ MeV.
 The $\chi^2$ values in the first model (column 1 in the Table~\ref{tab4})
 is too large and this model contradicts to the experimental data.
 The second and third models (columns 2 and 3 in Table~\ref{tab4}) are in
 agreement with the experimental data. The $\chi^2_{tot}$ value for the second
 model is less than for the third one. If $\sigma(\rho\to 3\pi)$,
 $\phi_{\omega\rho}$ and $r_0$ are the free parameters in a fit, then
 they are found to be equal to $\sigma(\rho\to 3\pi)=0.11\pm^{0.06}_{0.04}$
 nb,  $\phi_{\omega\rho}=-136\pm^{12}_{10}$ degree and $r_0=0.2\pm 0.3$
 GeV$^{-1}$. In this case the value of the range parameter $r_0$ turns out
 to be rather small
 and consistent with zero. The value of $\chi^2_{SND}$ for the second model is
 less than for the third one, and vice versa the $\chi^2_\omega$ for the third
 model is less than for the second. In the energy range
 $880\leq\sqrt[]{s}\leq 970$ MeV (Fig.\ref{crf4}) the fitted curves for the
 second and third models exceed the experimental points  in average by about 
 1.5 and 1 errors respectively. The difference between models 2 and 3 is also 
 seen in the
 energy regions $\sqrt[]{s}\leq 720$ MeV and $\sqrt[]{s}\geq 1100$ MeV
 (Fig.\ref{crf1} and \ref{crf6}). The $\chi^2_{\omega\pi\pi(DM2)}$ for the
 third model increases by a factor of 2 in comparison with the second model.
 The results of the $e^+e^-\to\omega\pi^+\pi^-$ cross section fits are shown in
 Fig.\ref{crsopp}. In case when the form factors (\ref{cform}) are used,
 the theoretical curve poorly describes experimental points
 at the left slope of the resonance . The CMD2 results of the 
 $e^+e^-\to\omega\pi^+\pi^-$
 reaction studies \cite{kmd2opp} are also presented in Fig.\ref{crsopp}.
 These data agree better with the second model.

 If the relative bias between SND and DM2 measurements is not assumed,
 the $\chi^2_{3\pi(DM2)}$ value is rather large:
 $\chi^2_{3\pi(DM2)}/N_{fit}=(37\div 40)/18$. Here $N_{fit}$ is the number of
 fitted experimental points (Table~\ref{tab4}). A rather large scale factor
 $C=1.54$ is required to concert the SND and DM2 data, and in this case
 $\chi^2_{3\pi(DM2)}/N_{fit}=(22\div 27)/18$.
 In order not to guess about relative systematics between the SND and DM2
 experiments, the fits described above were redone assuming $C=1$, but
 without taking into account $e^+e^-\to\pi^+\pi^-\pi^0$ cross section measured
 in DM2 experiments. (Table~\ref{tab5}). The parameters
 $m_{\omega^{\prime\prime}}$, $\Gamma_{\omega^{\prime\prime}}$ and
 $\sigma(\omega^{\prime\prime}\to\omega\pi^+\pi^-)$ were obtained from the
 fitting to the cross section $e^+e^-\to\omega\pi^+\pi^-$ reported by DM2, and
 $m_{\omega^\prime}$, $\Gamma_{\omega^\prime}$,
 $\sigma(\omega^\prime\to 3\pi)$, $\sigma(\omega^{\prime\prime}\to 3\pi)$ were
 obtained by using SND data only. In this case the first model (column 1,
 Table~\ref{tab5}) agrees with experimental data, but agreement is 
 significantly
 better if the fits are performed in the second or third models (columns 2 and
 3 in Table~\ref{tab5}). The $\chi^2_{tot}$ value for the third model is
 slightly bigger than for the second one. In this approach the fitted curve is
 in conflict with the DM2 measurenment of the $e^+e^-\to\pi^+\pi^-\pi^0$ cross
 section  (Fig.\ref{crf8}). 
 
\section{Discussion}

 Comparison of the $e^+e^-\to\pi^+\pi^-\pi^0$ cross section obtained in SND
 experiments with other results 
 \cite{cord,nd,ndnerez,kmd2omega,kmd2phi1,kmd2phi2} is shown in
 Fig.\ref{raz1},\ref{raz2},\ref{raz3} and \ref{raz4}. The DM1 results
 \cite{cord} are in agreement with the SND measurements. The ND results 
 \cite{nd,ndnerez}  agree with SND data in the energy region $\sqrt[]{s}< 930$
 MeV, while for $\sqrt[]{s}>930$ MeV ND points lay on about two errors lower
 than SND ones. In the vicinity of the $\omega$ resonance peak
 ($\sqrt[]{s}\simeq 780$ MeV) the SND cross section exceeds the CMD2 
 measurements \cite{kmd2omega}, while in the $\phi$ meson energy region the
 SND and CMD2 results \cite{kmd2phi1,kmd2phi2} are in agreement. 

 The $\omega$ meson parameters $m_\omega$, $\Gamma_\omega$,
 $\sigma(\omega\to 3\pi)$ were measured through study of the
 $e^+e^-\to\pi^+\pi^-\pi^0$ cross section.
 The $\omega$ meson mass was found to be
 $$ m_\omega = 782.79 \pm 0.08 \pm 0.09 \mbox{~~MeV}. $$
 Here the systematic error is related to the accuracy of the VEPP-2M energy 
 scale
 calibration by resonant depolarization method, 0.04 MeV, and to the model 
 uncertainty,
 0.08 MeV. The SND measurement in comparison with the results of experiments
 \cite{kmd2omega,spec95,cbar,olyaomega,cntr76,cord} and the world average
 value $m_\omega$ \cite{pdg} is shown in Fig.\ref{massa}. SND result is in
 agreement with the CMD2 measurement \cite{kmd2omega}, and differs from
 world average by about 1.3 standard deviations. The maximum difference, 
 about 3.4
 standard deviations is between the SND result and the Crystal Barrel 
 measurement $m_\omega=781.96 \pm 0.21$ \cite{cbar}.

 The following value of the $\omega$ meson width was obtained:
 $$ \Gamma_\omega = 8.68 \pm 0.04 \pm 0.15 \mbox{~~MeV}. $$
 The systematic error is related to the model dependence and to the accuracy 
 of energy
 determination. The comparison of this value with the results obtained in
 Ref.\cite{kmd2omega,spec95,ndomega,kmd,olyaomega,cord,benak}
 and with PDG world average value \cite{pdg} is shown in Fig.\ref{shira}.
 The SND result agrees with other measurements.

 The parameter $\sigma(\omega\to 3\pi)$ was found to be
 $$\sigma(\omega\to 3\pi) = 1615 \pm 9 \pm 57 \mbox{~~nb}.$$
 The systematic error includes the systematic uncertainties in the detection
 efficiency and luminosity determinations, 55 nb in total, and the model 
 dependence
 13 nb. The comparison of the obtained value with other experimental results
 \cite{kmd2omega,ndomega,kmd,olyaomega,cord,benak} and with PDG world average
 \cite{pdg} is shown in Fig.\ref{sech}. The SND result exceeds the
 central values of the majority of the  previous measurements. It differs by 
 less than one
 standard deviation from the results in Ref.\cite{ndomega,kmd,benak}, by about
 1.4 standard deviations from the DM1 measurement \cite{cord}, by 2 standard
 deviations from the OLYA result \cite{olyaomega} and PDG world average
 $\sigma(\omega\to 3\pi)=1484 \pm 29$ nb. The difference from the most precise
 measurements, done by CMD2 \cite{kmd2omega} and SND, is about 2.5
 standard deviations.

 Using the SND result $\sigma(\omega\to\pi^0\gamma)=155.8\pm 2.7\pm 4.8$ nb
 \cite{pi0gam}, the ratio of the partial widths of the $\omega\to\pi^0\gamma$ 
 and $\omega\to 3\pi$ decays  was calculated
 $$ {\Gamma(\omega\to\pi^0\gamma) \over \Gamma(\omega\to 3\pi)} = 0.097 \pm
    0.002 \pm 0.005 $$
 This value agrees with PDG world average \cite{pdg} and with other 
 experimental
 results \cite{sndot,ndomega,cntr76,benak,hlbc} (Fig.\ref{otnshir}).

 Using $\sigma(\omega\to 3\pi)$, measured in this work, the SND result of the
 $\omega\to\pi^0\gamma$ decay study \cite{pi0gam} and PDG world average value
 $B(\omega\to\pi^+\pi^-) = 0.0170\pm 0.0028$ \cite{pdg}, the partial width of
 the $\omega\to e^+e^-$ decay and the $\omega$ meson main decays branching 
 ratios were obtained:
 $$\Gamma(\omega\to e^+e^-) = 0.653 \pm 0.003 \pm 0.021 \mbox{~~keV},$$
 $$B(\omega\to e^+e^-) = (7.52 \pm 0.04 \pm 0.24) \times 10^{-5},$$
 $$B(\omega\to 3\pi) =  0.8965 \pm 0.0016 \pm 0.0048,$$
 $$B(\omega\to\pi^0\gamma) = 0.0865 \pm 0.0016 \pm 0.0042.$$
 Comparison of these results with PDG data \cite{pdg} is presented in 
 Table~.\ref{tab6}. The value of $B(\omega\to e^+e^-)$, calculated by using 
 SND data, exceeds the world average by about 2 standard deviations (by 8\%).
 
 The $e^+e^-\to\pi^+\pi^-\pi^0$ and $e^+e^-\to\omega\pi^+\pi^-$ cross sections
 analysis show that the data cannot be described by a sum of $\omega$, $\phi$
 mesons and two $\omega^\prime$, $\omega^{\prime\prime}$ resonances (model 1).
 The data can be satisfactory described with the model 3, which takes into 
 account form factors (\ref{cform}), with constrained partial widths growth 
 with energy. 
 The range parameter of this form factors was found to be
 $$ r_0 = 2.5 \pm^{1.1}_{0.8} \pm 0.5 \mbox{~~GeV}^{-1}.$$
 This agrees with expected vector meson effective ``radius'': 
 $2.5$--$3$ GeV$^{-1}$~
 \cite{pionuc}. The second error is due to model dependence.
 The model 2, which takes into account the  $\gamma^\star\to\rho\to 3\pi$
 transition, also satisfactory describes the experimental data. For parameters
 of this model, the following values were obtained:
 $$\sigma(\rho\to 3\pi)= 0.112 \pm^{0.060}_{0.040}\pm 0.038 \mbox{~~nb},$$
 $$\phi_{\omega\rho}=-135\pm^{17}_{13} \pm 9 \mbox{~~degree}.$$
 Here the systematic error is due to model uncertainty. The above given
 $\sigma(\rho\to 3\pi)$ value corresponds to the branching ratio
 $B(\rho\to 3\pi)=(1.01\pm^{0.54}_{0.36}\pm 0.34)\times 10^{-4}$.
 Assuming that the $\rho\to 3\pi$ transition proceeds via the $\rho-\omega$
 mixing mechanism, the following values of the $\gamma^\star\to\rho\to 3\pi$
 process parameters are expected: $\phi_{\omega\rho}\simeq -90^\circ$ and
 $\sigma(\rho\to 3\pi) = 0.05$ -- $0.07$ nb. The $\sigma(\rho\to 3\pi)$ value
 obtained in the analysis agrees with the expected one, while 
 $\phi_{\omega\rho}$
 differs from the expected value by about two standard deviations.

 In general, the model 2 seems to be more preferable than the model 3 due to 
 the following
 considerations. The full data set for the $e^+e^-\to \pi^+\pi^-\pi^0$ and
 $\omega\pi^+\pi^-$ cross sections is in somewhat better agreement with the
 second model. Approximation of the $e^+e^-\to\omega\pi^+\pi^-$ cross section 
 with the third model is poor (Fig.\ref{crsopp}). The $\phi_{\omega\phi}$ phase
 value, obtained by the fit in the second model agrees with the theoretical
 prediction $\phi_{\omega\phi}=\Psi(m_\phi)\approx 160^\circ$ \cite{faza},
 while the phase $\phi_{\omega\phi}$, obtained by using the third model,
 exceeds the expected
 value (Fig.\ref{faza}). But, unfortunately, available experimental data are
 insufficient to make a strict conclusion about observation of the 
 $\rho\to 3\pi$ decay.
 
 The parameter $\sigma(\phi\to 3\pi)$ was found to be
 $$\sigma(\phi\to 3\pi) = 657 \pm 10 \pm 37 \mbox{~~nb}.$$
 The systematic error includes the systematic uncertainties in the detection
 efficiency and luminosity determinations, 33 nb in total, and the model 
 dependence 17 nb. This value agrees with the results of our previous analysis
 \cite{phi98,pi3mhad}.
 The SND result agrees also with other measurements
 \cite{kmd2phi2,kmd2phi1,nd,olyaphiomega,cord,buk,parr,cosm} and with
 PDG world average \cite{pdg2000} (Fig.\ref{phisech}).

 The fit within the models 2 and 1 (Tables \ref{tab4} and \ref{tab5}) gave the
 result
 $$\phi_{\omega\phi}=163 \pm 3 \pm 6 \mbox{~~degree}.$$
 The systematic error is related to model dependence.
 The obtained result is in agreement with the theoretical prediction
 (Fig.\ref{faza}) $\phi_{\omega\phi}=\Psi(s), \Psi(m_\phi)=163^\circ$, which
 takes into account the $\phi-\omega$ mixing \cite{faza}. The fits with the
 model 3 (column 3 in Tables~\ref{tab4} and \ref{tab5}) gave the result
 $\phi_{\omega\phi}=190\pm 5 \pm 10$ degree, which exceeds the theoretical
 prediction.

 The conventional view on the $OZI$ suppressed  $\phi\to\pi^+\pi^-\pi^0$ decay
 is that it proceeds through $\phi-\omega$ mixing, i.e.
 in the wave function of the $\phi$-meson, which is dominated by $s$ quarks,
 there is an admixture of $u$ and $d$ quarks:
 $$|\phi>\approx |\phi^{(0)}> + \varepsilon_{\phi\omega}|\omega^{(0)}>,
 \mbox{~~} |\phi^{(0)}>=s\overline{s},\mbox{~~}
 |\omega^{(0)}>=(u\overline{u}+d\overline{d})/\sqrt[]{2},$$
 $\varepsilon_{\phi\omega}\approx 0.05$ is $\phi-\omega$ mixing parameter.
 An alternative to the $\phi-\omega$ mixing is the direct decay. In
 Ref.\cite{akphiom1,freshlook,akozi,akphiom2,akphiom3} it was shown that there
 are no serious reasons to prefer the $\phi-\omega$ mixing to the direct
 transition, and methods of determination of the $\phi\to\pi^+\pi^-\pi^0$
 decay mechanism were suggested.
 In particular it was proposed in Ref.\cite{akphiom3} to analyze the
 $\Gamma(\phi\to e^+e^-)/\Gamma(\omega\to e^+e^-)$ ratio. In this work the
 $B(\omega\to e^+e^-)$ based mainly on the SND data was obtained, and in the 
 Ref.\cite{phi98} the $B(\phi\to e^+e^-)$  was measured by SND. We performed 
 the analysis of the $\phi$ and $\omega$ mesons lepton widths ratio based on 
 the SND data only.
 To improve the accuracy of $B(\phi\to e^+e^-)$ determination, the SND results
 of the $\phi\to\mu^+\mu^-$ decay studies \cite{sndmumu} were used. The 
 average of these measurements
 $\sqrt[]{B(\phi\to e^+e^-)B(\phi\to\mu^+\mu^-)}=(2.93\pm 0.11)\times 10^{-4}$
 agrees with $B(\phi\to e^+e^-)=(2.93\pm 0.15)\times 10^{-4}$ ~\cite{phi98}.
 Assuming  $B(\phi\to e^+e^-)=B(\phi\to\mu^+\mu^-)$, one gets
 $B(\phi\to e^+e^-)=(2.93\pm 0.09)\times 10^{-4}$. The ratio of the leptonic 
 widths is equal to
 $$ R_{e^+e^-}={\Gamma(\phi\to e^+e^-) \over \Gamma(\omega\to e^+e^-)}=
 1.89\pm 0.08$$
 On the other hand this ratio can be written in the following form:
\begin{eqnarray}
 \label{rmix}
 R_{e^+e^-}= \Biggl({m_\omega \over m_\phi}\Biggr)^3
 \Biggl|{ {g_{\gamma\phi}^{(0)}+\varepsilon_{\phi\omega}g_{\gamma\omega}^{(0)}}
 \over {g_{\gamma\omega}^{(0)}-\varepsilon_{\phi\omega}g_{\gamma\phi}^{(0)}}
 }\Biggr|^2.
\end{eqnarray}
 Using the obtained $R_{e^+e^-}$ value, equation (\ref{rmix}) and
 the nonrelativistic quark model prediction:
 $${f_\omega^{(0)} \over f_\phi^{(0)}}= - \sqrt[]{2} \mbox{~~~}
 \biggr(f_V = {\sqrt[]{\alpha} m_V^2 \over g_{\gamma V} } \biggl),$$
 the $\phi-\omega$ mixing parameter $\varepsilon_{\phi\omega}\approx 0.06$ was
 obtained. On the other hand, taking into account the equation
 $g_{\phi\rho\pi} = g_{\phi\rho\pi}^{(0)}+ \varepsilon_{\phi\omega}
 g_{\omega\rho\pi}^{(0)}$,
 and assuming $g_{\phi\rho\pi}^{(0)} = 0$ and
 $g_{\omega\rho\pi}^{(0)}=g_{\omega\rho\pi}$, we found
 $\varepsilon_{\phi\omega}\approx 0.06$. Here $g_{\omega\rho\pi}$ and
 $g_{\phi\rho\pi}$ coupling constants were calculated by using the SND results
 obtained in this work and in Ref.\cite{phi98}, the phase space factors
 $W_{\rho\pi}(m_\omega)$ and $W_{\rho\pi}(m_\phi)$ were calculated assuming 
 $s_1=0$ and $a_{3\pi}=0$. In this case the SND data agree with the 
 $\phi-\omega$ mixing dominance in the $\phi\to\pi^+\pi^-\pi^0$ decay.
 
 The ratio $f_\omega^{(0)}/f_\phi^{(0)}=-\sqrt[]{2}$ is valid if 
 $\psi(0,m_V)$ wave function of the $q\overline{q}$ bound state at the origin
 behaves like $|\psi(0,m_V)|^2\propto m_V^3$, that is corresponds to the 
 Coulomb-like
 nonrelativistic potential. But experimental data on the vector mesons $\rho$,
 $\omega$, $\phi$, $J/\psi$, $\Upsilon(1S)$ leptonic widths support the
 $|\psi(0,m_V)|^2\propto m_V^2$ behavior. Indeed, according to
 Ref.\cite{shiree}:
 $$
 \Gamma(V\to e^+e^-) = {16\pi\alpha^2\over m_V^2} C_V^2 |\psi(0,m_V)|^2,
 $$
 where $C_V$ is the mean electric charge of the valence quarks inside the
 vector meson $V$ ($C_\rho^2=1/2$, $C_\omega^2=1/18$, $C_\phi^2=1/9$,
 $C_{J/\psi}^2=4/9$, $C_{\Upsilon(1S)}^2=1/9$).
 In the case of $|\psi(0,m_V)|^2\propto m_V^2$, and in the absence of the 
 mixing, the following ratios are expected:
\begin{eqnarray}
\label{otnshiree}
\nonumber 
{ \Gamma(\rho\to e^+e^-):\Gamma(\omega\to e^+e^-):\Gamma(\phi\to e^+e^-):
 \Gamma(J/\psi\to e^+e^-):\Gamma(\Upsilon(1S)\to e^+e^-) =} \\
 = C_\rho^2: C_\omega^2: C_\phi^2: C_{J/\psi}^2: C_{\Upsilon(1S)}^2=
 4.5:0.5:1:4:1
\end{eqnarray}
 Using the SND $\Gamma(\omega\to e^+e^-)$ and $\Gamma(\phi\to e^+e^-)$ values 
 and the world average for the $\Gamma(\rho\to e^+e^-)$, 
 $\Gamma(J/\psi\to e^+e^-)$ and
 $\Gamma(\Upsilon(1S)\to e^+e^-)$  \cite{pdg}, we have found:
\begin{eqnarray}
\nonumber 
{  \Gamma(\rho\to e^+e^-):\Gamma(\omega\to e^+e^-):\Gamma(\phi\to e^+e^-):  
  \Gamma(J/\psi\to e^+e^-):\Gamma(\Upsilon(1S)\to e^+e^-) = } \\
  = 5.2 \pm 0.2: 0.495 \pm 0.025 : 0.93 \pm 0.05 : 3.98 \pm 0.032 : 1,
\end{eqnarray}
 This ratios agree with the expected (\ref{otnshiree}).
 If $|\psi(0,m_V)|^2\propto m_V^2$, then
 $f_\omega^{(0)} / f_\phi^{(0)} = - \sqrt[]{2m_\omega/m_\phi}$
 and by using Eq.(\ref{rmix}) $\varepsilon_{\phi\omega}\approx 0.015$ can be 
 obtained. In this case the coupling constant of the
 direct $\phi\to\pi^+\pi^-\pi^0$ decay 
 $g_{\phi\rho\pi}^{(0)} \approx 0.7 \cdot g_{\phi\rho\pi}$ is required to
 describe the experimental value of $B(\phi\to 3\pi)$, indicating the direct
 transition as the main mechanism of the decay.

 The following $\omega^\prime$ parameters were obtained from the fits 
 (Table \ref{tab4} and \ref{tab5}):
$$
 m_{\omega^\prime} = 1400 \pm 50 \pm 130  \mbox{~~MeV},
$$
$$
 \Gamma_{\omega^\prime} = 870 \pm^{500}_{300} \pm 450 \mbox{~~MeV},
$$
$$
 \sigma(\omega^\prime\to 3\pi) = 4.9 \pm 1.0 \pm 1.6 \mbox{~~nb}.
$$
 The $\omega^\prime$ decays mostly into $\pi^+\pi^-\pi^0$ and its electronic
 width is $\Gamma(\omega^\prime\to e^+e^-) \sim 570$ eV. The
 $\omega^{\prime\prime}$ parameters were found to be
$$
 m_{\omega^{\prime\prime}} = 1770 \pm 50 \pm 60 \mbox{~~MeV},
$$
$$
 \Gamma_{\omega^{\prime\prime}} = 490 \pm^{200}_{150} \pm 130 \mbox{~~MeV},
$$
$$
 \sigma(\omega^{\prime\prime}\to 3\pi) = 5.4 \pm^{2.0}_{0.4} \pm 3.9 
 \mbox{~~nb},
$$
$$
 \sigma(\omega^{\prime\prime}\to\omega\pi^+\pi^-) = 1.9 \pm 0.4 \pm 0.6 
 \mbox{~~nb}.
$$
 The $\omega^{\prime\prime}$ resonance decays with approximately equal
 probability into $\pi^+\pi^-\pi^0$ and $\omega\pi\pi$:
 $B(\omega^{\prime\prime}\to 3\pi)\sim 0.65$,
 $B(\omega^{\prime\prime}\to \omega\pi\pi) \sim 0.35$ and it has the
 electronic width  $\Gamma(\omega^{\prime\prime}\to e^+e^-) \sim 860$ eV.
 The second errors shown above are due to the model uncertainty and possible
 bias between the SND and DM2 measurements. The $\omega^\prime$ and
 $\omega^{\prime\prime}$ parameters, obtained in this work, are somewhat
 different from those obtained in our previous analysis \cite{pi3mhad}.
 In particular, the $\omega^\prime$ and $\omega^{\prime\prime}$ full widths
 values decrease a little. This difference is attributed to the fact that
 a new data for the $e^+e^-\to\pi^+\pi^-\pi^0$ cross section below 1 GeV were
 added in the fits. Of cause, the obtained values are not precise 
 measurements, they should be considered as rather approximate estimation of
 the $\omega^\prime$ and $\omega^{\prime\prime}$ resonances main parameters.
 
 In the energy region  $880 \leq \sqrt[]{s} \leq 970$ MeV, experimental
 points deviate from fitting curves (Fig.\ref{crf4}). The difference can be 
 attributed
 to inadequacy of the applied theoretical models, uncertainty of the 
 $\omega^\prime$
 and $\omega^{\prime\prime}$ resonances contributions. Maybe more accurate
 consideration of the vector mesons mixing is required.

 Analysis of the 

 Using the $e^+e^-\to\pi^+\pi^-\pi^0$ cross section, obtained with SND 
 detector in
 this work and in Ref.\cite{phi98,pi3mhad}, the contribution to the anomalous
 magnetic moment of the muon, due to the $\pi^+\pi^-\pi^0$ intermediate state 
 in the vacuum polarization, was calculated via dispersion integral:
$$ a_\mu(3\pi, \sqrt[]{s}<1.38\mbox{GeV})=\biggl({\alpha m_\mu \over 3\pi}
   \biggr)^2 \int^{s_{max}}_{s_{min}} {R(s)K(s) \over s^2} ds,
$$
 where $s_{max}=1.38$ GeV, $s_{min}=m_{\pi^0}+2m_\pi$, $K(s)$ is the QED 
kernel,
$$
 R(s) = {\sigma(e^+e^-\to\pi^+\pi^-\pi^0) \cdot 
  (1-\Delta_l(s)-\Delta_h(s))^2 \over \sigma(e^+e^-\to\mu^+\mu^-)}, \mbox{~~}
  \sigma(e^+e^-\to\mu^+\mu^-) = {4 \pi \alpha \over 3 s}.
$$
 Here $\sigma(e^+e^-\to\pi^+\pi^-\pi^0)$ is the experimental cross section,
 $\Delta_l(s)$ and $\Delta_h(s)$ are corrections due to leptonic and hadronic
 vacuum polarizations. The $\Delta_l(s)$ was calculated according to 
 Ref.\cite{arbuz}
 and  $\Delta_h(s)$ was obtained by using $e^+e^-\to hadrons$ total cross 
 section.

 The integral was evaluated by using the trapezoidal rule. To take into account
 the numerical integration errors, the correction method suggested in
 Ref.\cite{aki} was applied. As a result we obtained:
 $$a_\mu(3\pi, \sqrt[]{s}<1.38\mbox{GeV}) = (458 \pm 2 \pm 17) \times 10^{-11}.$$

 At present in BINP(Novosibirsk) the VEPP-2000 collider with the energy range 
 from 0.36 to 2 GeV and luminosity up to $10^{32}$ cm$^{-2}$s$^{-1}$ (at
 $\sqrt[]{s} \sim 2$ GeV) is under construction \cite{vepp2000}.
 The $e^+e^-\to \pi^+\pi^-\pi^0$ process studies in the
 energy region $\sqrt[]{s}<2$ GeV will be continued in future experiments with 
 SND detector at this new facility. 

\section{Conclusion}
 The cross section of the process $e^+e^-\to \pi^+\pi^-\pi^0$ was measured in
 the SND experiment at the VEPP-2M collider in the energy region
 $\sqrt[]{s}$ below 980 MeV.
 The measured cross section was analyzed in the framework of the generalized
 vector meson dominance model together with the $e^+e^-\to \pi^+\pi^-\pi^0$
 and $\omega\pi^+\pi^-$ cross sections obtained by SND and DM2 in the energy
 region $980 < \sqrt[]{s} < 2000$ MeV.
 The $\omega$-meson parameters: $m_\omega=782.79\pm 0.08\pm 0.09$ MeV,
 $\Gamma_\omega=8.68\pm 0.04\pm 0.15$ MeV and
 $\sigma(\omega\to 3\pi)=1615\pm 9\pm 57$ nb were obtained.

 It was found that
 the experimental data cannot be described by a sum of $\omega$, $\phi$,
 $\omega^\prime$ and $\omega^{\prime\prime}$ resonances contributions. This
 can be interpreted as manifestation of $\rho\to 3\pi$ decay suppressed by 
 $G$-parity, with relative probability
 $B(\rho\to 3\pi) = (1.01\pm^{0.54}_{0.36}\pm 0.034) \times 10^{-4}$.
 The relative interference phase between the $\omega$ and $\rho$ mesons was
 found to be equal to: $\phi_{\omega\rho} = -135 \pm^{17}_{13}\pm 9$ degree.
 These parameters of the $\rho\to 3\pi$ decay are in agreement with the 
 theoretical values expected from the $\rho-\omega$ mixing.

 Analysis of the $\Gamma(\phi\to e^+e^-)/\Gamma(\omega\to e^+e^-)$ ratio
 and $g_{\phi\rho\pi}$ and $g_{\omega\rho\pi}$ coupling constants obtained in
 SND experiments indicates that the direct transition is preferable to
 $\phi-\omega$ mixing as the main mechanism of the $\phi\to\pi^+\pi^-\pi^0$
 decay.

 Using the $e^+e^-\to\pi^+\pi^-\pi^0$ cross section, obtained with SND 
 detector the contribution to the anomalous magnetic moment of the muon, due
 to the $\pi^+\pi^-\pi^0$ intermediate state  in the vacuum polarization,
 was calculated:  
 $a_\mu(3\pi,\sqrt[]{s}<1.38\mbox{GeV})=(458 \pm 2 \pm 17) \times 10^{-11}.$

\section*{acknowledgments}

 The authors are grateful to N.N.Achasov for useful discussions. The present
 work was supported in part by grant no. 78 1999 of Russian Academy of Science
 for young scientists and by Russian Science Support Foundation.

\newpage

\begin{table}[!hbp]
\begin{center}
\caption{Event numbers $N_{3\pi}$ of the $e^+e^-\to\pi^+\pi^-\pi^0(\gamma)$
process (after background subtraction) and $N_{bkg}$ of background processes,
integrated luminosity $IL$ and detection efficiency $\epsilon(s,E_\gamma=0)$
(without $\gamma$-quantum radiation). $\delta_{rad}$ is the radiative 
correction [$\delta_{rad}=\xi(s)/\epsilon(s,E_\gamma=0)$, $\xi(s)$ is defined
via the expression (\ref{xifu})].}
\label{tab1}
\begin{tabular}[t]{cccccc}
$\sqrt[]{s}$ (MeV)&$IL$ (nb$^{-1}$)&$\epsilon(s,E_\gamma=0)$&$N_{3\pi}$ &
$N_{bkg}$&$\delta_{rad}$ \\ \hline
970&271.4$\pm$2.7&0.2544$\pm$0.0082&   800$\pm$ 34&33$\pm$7&0.905 \\
958&249.0$\pm$2.5&0.2544$\pm$0.0082&   658$\pm$ 29&27$\pm$5&0.918 \\
950&276.8$\pm$2.7&0.2540$\pm$0.0086&   727$\pm$ 32&32$\pm$6&0.927 \\
940&505.2$\pm$4.5&0.2585$\pm$0.0075&  1203$\pm$ 41&53$\pm$10&0.937 \\
920&510.1$\pm$4.1&0.2699$\pm$0.0075&  1292$\pm$ 42&52$\pm$ 8&0.96 \\
880&397.6$\pm$3.6&0.3268$\pm$0.0032&  1596$\pm$ 49&61$\pm$19&1.094 \\
840&711.0$\pm$6.1&0.3341$\pm$0.0029&  5928$\pm$ 88&96$\pm$15&1.356 \\
820&329.0$\pm$3.0&0.3376$\pm$0.0022&  5478$\pm$ 84&74$\pm$13&1.491 \\
810.40&223.7$\pm$2.1&0.3384$\pm$0.0019&  5989$\pm$ 86&27$\pm$ 5&1.463 \\
809.79& 67.8$\pm$0.8&0.3412$\pm$0.0012&  1899$\pm$ 48& 5$\pm$ 1&1.464 \\
800.40&235.6$\pm$2.2&0.3399$\pm$0.0014& 11694$\pm$121&40$\pm$10&1.319 \\
799.79& 53.6$\pm$0.7&0.3435$\pm$0.0013&  2679$\pm$ 57& 6$\pm$ 2&1.308 \\
794.40&160.8$\pm$1.6&0.3408$\pm$0.0012& 13757$\pm$129&19$\pm$ 4&1.165 \\
793.79& 54.8$\pm$0.7&0.3448$\pm$0.0010&  5066$\pm$ 78& 8$\pm$ 1&1.148 \\
790.40&136.3$\pm$1.4&0.3414$\pm$0.0011& 20228$\pm$157&15$\pm$ 4&1.036 \\
789.79& 58.8$\pm$0.7&0.3458$\pm$0.0010&  9054$\pm$104& 7$\pm$ 1&1.015 \\
786.40&177.6$\pm$1.7&0.3420$\pm$0.0010& 51265$\pm$251&27$\pm$ 5&0.895 \\
786.18&20.4$\pm$0.4&0.3450$\pm$0.0037&6226$\pm$ 88& 3$\pm$ 1&0.887 \\
785.79& 76.9$\pm$0.9&0.3466$\pm$0.0010& 24876$\pm$175&10$\pm$ 1&0.874 \\
785.40&222.4$\pm$2.1&0.3422$\pm$0.0010& 75531$\pm$304&33$\pm$ 7&0.861 \\
784.40&285.3$\pm$2.7&0.3424$\pm$0.0010&111828$\pm$371&34$\pm$ 5&0.830 \\
783.79& 78.1$\pm$0.9&0.3470$\pm$0.0010& 33325$\pm$201& 7$\pm$ 3&0.814 \\
783.40&288.5$\pm$2.6&0.3424$\pm$0.0010&122114$\pm$387&80$\pm$10&0.804 \\
782.90&122.3$\pm$1.2&0.3477$\pm$0.0012& 54830$\pm$261&40$\pm$ 7&0.794 \\
782.79& 85.2$\pm$0.9&0.3473$\pm$0.0010& 37956$\pm$217&16$\pm$ 3&0.792 \\
782.40&300.9$\pm$2.7&0.3426$\pm$0.0010&127682$\pm$397&36$\pm$14&0.785 \\
782.13&15.1$\pm$0.3&0.3534$\pm$0.0037&6452$\pm$ 89& 4$\pm$ 2&0.781 \\
781.79&372.5$\pm$3.3&0.3475$\pm$0.0010&155515$\pm$436&49$\pm$17&0.777 \\
781.40&220.4$\pm$2.1&0.3427$\pm$0.0010& 85611$\pm$324&85$\pm$10&0.773 \\
780.40&169.2$\pm$1.6&0.3429$\pm$0.0010& 56031$\pm$262&10$\pm$ 1&0.767 \\
778.11&20.9$\pm$0.4&0.3534$\pm$0.0031&4344$\pm$ 72& 8$\pm$ 4&0.767 \\
780.79&131.9$\pm$1.3&0.3477$\pm$0.0010& 48230$\pm$241&42$\pm$ 6&0.769 \\
779.90&114.7$\pm$1.2&0.3457$\pm$0.0014& 34860$\pm$207&14$\pm$ 5&0.766 \\
779.79& 44.7$\pm$0.6&0.3478$\pm$0.0010& 13099$\pm$126& 1$\pm$ 1&0.766 \\
778.40&159.6$\pm$1.6&0.3432$\pm$0.0011& 34568$\pm$207&21$\pm$ 3&0.767 \\
777.79& 79.2$\pm$0.9&0.3483$\pm$0.0010& 14700$\pm$134&10$\pm$ 4&0.768 \\
774.40&162.2$\pm$1.6&0.3439$\pm$0.0012& 14157$\pm$131&21$\pm$ 6&0.779 \\
773.79& 65.1$\pm$0.8&0.3492$\pm$0.0010&  4952$\pm$ 78&10$\pm$ 1&0.781 \\
770.40&253.5$\pm$2.3&0.3445$\pm$0.0013& 10959$\pm$116&33$\pm$ 7&0.792 \\
769.79& 45.9$\pm$0.6&0.3500$\pm$0.0011&  1656$\pm$ 44&10$\pm$ 1&0.794 \\
764.40&222.8$\pm$2.1&0.3455$\pm$0.0015&  4242$\pm$ 71&31$\pm$ 7&0.806 \\
763.79& 40.2$\pm$0.6&0.3512$\pm$0.0013&   724$\pm$ 30& 5$\pm$ 1&0.808 \\
760.40&208.2$\pm$2.0&0.3461$\pm$0.0017&  2658$\pm$ 57&19$\pm$ 6&0.814 \\
759.79& 43.5$\pm$0.6&0.3520$\pm$0.0014&   576$\pm$ 28& 7$\pm$ 3&0.815 \\
750.40&174.6$\pm$1.7&0.3479$\pm$0.0021&  1008$\pm$ 37&26$\pm$ 7&0.826 \\
749.79& 52.2$\pm$0.7&0.3541$\pm$0.0018&   251$\pm$ 18&14$\pm$ 5&0.828 \\
720&584.1$\pm$5.0&0.3563$\pm$0.0069&   652$\pm$ 30&60$\pm$ 8&0.848 \\
690&174.4$\pm$1.6&0.3526$\pm$0.0069&    58$\pm$ 11&21$\pm$ 5&0.860 \\
660&281.1$\pm$2.5&0.3575$\pm$0.0070&    40$\pm$ 11&29$\pm$ 4&0.862 \\
630&120.1$\pm$1.2&0.3532$\pm$0.0068&     0$\pm$  5&14$\pm$ 3&0.865 \\
600& 90.6$\pm$0.9&0.3298$\pm$0.0066&    -2$\pm$  6&15$\pm$ 4&0.868 \\
580& 12.7$\pm$0.2&0.3561$\pm$0.0069&     2$\pm$  4& 2$\pm$ 1&0.867 \\
560& 11.2$\pm$0.2&0.3369$\pm$0.0068&    -1$\pm$  1& 1$\pm$ 1&0.867 \\
540& 12.1$\pm$0.2&0.3156$\pm$0.0067&    -4$\pm$  2& 4$\pm$ 2&0.867 \\
520&  7.2$\pm$0.2&0.2866$\pm$0.0065&     0        & 0       &0.861 \\
500&  8.0$\pm$0.2&0.2278$\pm$0.0060&     0$\pm$  1& 1$\pm$ 1&0.856 \\
480& 13.4$\pm$0.2&0.2030$\pm$0.0058&     0        & 0       &0.852 \\
440&  6.2$\pm$0.1&0.0183$\pm$0.0019&     0        & 0       &0.820 \\
\hline
\end{tabular}
\end{center}
\end{table}

\begin{table}[h]
\begin{center}
\caption{The ratio of the cross sections, obtained by using different methods 
of background subtraction.
$N_{3\pi}$ is the number of the $e^+e^-\to 3\pi$ events obtained by fitting the
two-photon invariant mass spectra. $\sigma{(1)}$ -- the cross section measured
in the approach when background was calculated according to Eq.(\ref{bg}), 
$\sigma^{(2)}$ -- the cross section measured by using the two-photon invariant 
mass spectra analysis.}

\label{tab2}
\begin{tabular}[t]{cccccc}
$\sqrt[]{s}$ (MeV)&$N_{3\pi}$ &$\sigma^{(1)}/\sigma^{(2)}$ \\ \hline
750 & 1350$\pm$42 &0.995$\pm$0.034 \\
720 &  700$\pm$32 &0.999$\pm$0.054 \\
690 &   66$\pm$11 &0.956$\pm$0.209 \\
660 &   17$\pm$10 &2.562$\pm$1.114 \\
\hline
\end{tabular}
\end{center}
\end{table}

\begin{table}
\caption{The $e^+e^-\to\pi^+\pi^-\pi^0$ cross section.
$\sigma_{mod}$ is the model uncertainty, $\sigma_{bkg}$ is the error due to
background subtraction, $\sigma_{eff}\oplus\sigma_{IL}$ -- the error due to
uncertainty in the detection efficiency and integrated luminosity 
determination 
(3.4\% at $\sqrt[]{s}<900$ MeV and 4.5\% for $\sqrt[]{s}>900$ MeV),
$\sigma_{sys}=\sigma_{eff} \oplus \sigma_{IL} \oplus \sigma_{mod}(s) \oplus 
\sigma_{bkg}(s)$
is the total systematic error. }
\label{tab3}
\begin{center}
\small
\begin{tabular}[t]{cccccc}
$\sqrt[]{s}$(MeV)&$\sigma$(nb)&$\sigma_{bkg}$(nb)&$\sigma_{mod}$(nb)&
$\sigma_{eff}\oplus\sigma_{IL}$(nb)&$\sigma_{sys}$(nb)\\ \hline
970.00&  12.82$\pm$ 0.70&0.32&0.05& 0.58& 0.66 \\
958.00&  11.33$\pm$ 0.64&0.28&0.05& 0.51& 0.58 \\
950.00&  11.17$\pm$ 0.62&0.29&0.06& 0.50& 0.58 \\
940.00&   9.83$\pm$ 0.41&0.26&0.05& 0.44& 0.52 \\
920.00&   9.82$\pm$ 0.39&0.24&0.04& 0.44& 0.50 \\
880.00&  11.22$\pm$ 0.36&0.26&0.03& 0.38& 0.46 \\
840.00&  18.77$\pm$ 0.33&0.18&0.21& 0.64& 0.70 \\
820.00&  32.93$\pm$ 0.58&0.27&0.34& 1.12& 1.20 \\
810.40&  53.84$\pm$ 1.01&0.15&0.41& 1.83& 1.88 \\
809.79&  55.81$\pm$ 1.60&0.09&0.42& 1.90& 1.95 \\
800.40& 110.42$\pm$ 1.82&0.23&0.42& 3.75& 3.78 \\
799.79& 111.09$\pm$ 2.98&0.15&0.41& 3.78& 3.80 \\
794.40& 215.25$\pm$ 3.91&0.18&0.31& 7.32& 7.33 \\
793.79& 233.54$\pm$ 5.48&0.22&0.32& 7.94& 7.95 \\
790.40& 419.31$\pm$ 8.24&0.19&0.56&14.26&14.27 \\
789.79& 437.91$\pm$10.55&0.20&0.37&14.89&14.89 \\
786.40& 943.63$\pm$19.81&0.30&1.45&32.08&32.12 \\
786.18& 998.92$\pm$24.90&0.29&0.61&33.96&33.97 \\
785.79&1068.39$\pm$23.36&0.26&0.86&36.33&36.34 \\
785.40&1154.29$\pm$22.45&0.30&1.91&39.25&39.29 \\
784.40&1382.69$\pm$21.80&0.25&2.44&47.01&47.08 \\
783.79&1514.97$\pm$22.99&0.19&0.74&51.51&51.51 \\
783.40&1542.58$\pm$17.29&0.61&2.16&52.45&52.50 \\
782.90&1627.56$\pm$18.88&0.71&0.32&55.34&55.34 \\
782.79&1624.77$\pm$20.43&0.41&0.47&55.24&55.25 \\
782.40&1584.21$\pm$17.63&0.27&2.59&53.86&53.93 \\
782.13&1552.85$\pm$40.68&0.58&0.46&52.80&52.80 \\
781.79&1550.80$\pm$21.87&0.29&0.44&52.73&52.73 \\
781.40&1470.94$\pm$25.52&0.88&2.49&50.01&50.08 \\
780.79&1369.11$\pm$28.77&0.72&0.51&46.55&46.56 \\
780.40&1261.06$\pm$28.76&0.14&2.18&42.88&42.93 \\
779.90&1146.89$\pm$28.79&0.28&0.43&38.99&39.00 \\
779.79&1098.85$\pm$30.58&0.05&0.40&37.36&37.36 \\
778.40& 822.81$\pm$21.45&0.30&1.52&27.98&28.02 \\
778.11& 765.41$\pm$20.01&0.85&0.23&26.02&26.04 \\
777.79& 693.08$\pm$19.36&0.28&0.23&23.56&23.57 \\
774.40& 325.90$\pm$ 7.86&0.29&0.69&11.08&11.11 \\
773.79& 278.72$\pm$ 7.87&0.34&0.08& 9.48& 9.48 \\
770.40& 158.58$\pm$ 3.36&0.29&0.37& 5.39& 5.41 \\
769.79& 129.74$\pm$ 4.55&0.47&0.06& 4.41& 4.44 \\
764.40&  68.48$\pm$ 1.58&0.30&0.26& 2.33& 2.36 \\
763.79&  63.43$\pm$ 2.87&0.26&0.06& 2.16& 2.17 \\
760.40&  45.38$\pm$ 1.18&0.19&0.13& 1.54& 1.56 \\
759.79&  46.14$\pm$ 2.34&0.34&0.06& 1.57& 1.61 \\
750.40&  20.12$\pm$ 0.78&0.31&0.08& 0.68& 0.76 \\
749.79&  16.40$\pm$ 1.22&0.55&0.04& 0.56& 0.78 \\
720.00&   3.69$\pm$ 0.19&0.20&0.01& 0.13& 0.24 \\
690.00&   1.10$\pm$ 0.21&0.24&0.00& 0.04& 0.24 \\
660.00&   0.46$\pm$ 0.13&0.20&0.00& 0.02& 0.20 \\
630.0 &$<0.34$ (90\% CL)  &&&& \\
600.0 &$<0.53$ (90\% CL)  &&&& \\
580.0 &$<1.36$ (90\% CL)  &&&& \\
560.0 &$<0.44$ (90\% CL)  &&&& \\
540.0 &$<1.21$ (90\% CL)  &&&& \\
520.0 &$<1.3$ (90\% CL)  &&&& \\
500.0 &$<0.96$ (90\% CL)  &&&& \\
480.0 &$<0.99$ (90\% CL)  &&&& \\
440.0 &$<24.7$ (90\% CL)  &&&& \\
\end{tabular}
\end{center} 
\end{table}

\begin{table}[!hbp]
\caption{Fit results for the $e^+e^-\to\pi^+\pi^-\pi^0$ and $\omega\pi^+\pi^-$
         cross sections. The column number $N$ corresponds to the different
	 models for the $A_{\rho\pi}$ amplitude. $N_{fit}$ is the number of 
         fitted
	 points. The first error is statistical, the second error shows the
	 difference in the fit results due to various assumptions about 
	 the $e^+e^-\to\pi^+\pi^-\pi^0$ reaction dynamics and relative 
         systematics
	 between the SND and DM2 measurements.}
\label{tab4}
\begin{center}
\begin{tabular}[t]{llll}
 $N$&1&2&3 \\ \hline
$m_\omega,$ MeV&782.75$\pm$0.08&782.72$\pm$0.08&782.71$\pm$0.08 \\
$\Gamma_\omega,$ MeV&8.60$\pm$0.04$\pm$0.01&8.73$\pm$0.04$\pm$0.02&8.63$\pm
$0.04 \\
$\sigma(\omega\to 3\pi),$ nb &1609$\pm$7$\pm$1&1624$\pm$10$\pm$5&1610$\pm$7
$\pm$3 \\
$\sigma(\phi\to 3\pi)$, nb&645$\pm$6$\pm$4&645$\pm$7$\pm$6&658$\pm$8$\pm$7 \\
$\phi_{\omega\phi}$, degree&161$\pm$2$\pm$4&163$\pm^3_2\pm$4&187$\pm$4$\pm$6 \\
$r_0$, GeV$^{-1}$&&&2.6$\pm^{1.1}_{0.8}\pm$0.2 \\
$m_{\omega^\prime}$, MeV&
1358$\pm$20$\pm$45&1460$\pm^{70}_{50}\pm$70&1410$\pm$30$\pm$60 \\
$\Gamma_{\omega^\prime}$, MeV&
500$\pm^{60}_{50}\pm$80&1120$\pm^{500}_{300}\pm$200&617$\pm$40$\pm$95 \\
$\sigma(\omega^\prime\to 3\pi)$, nb&
5.7$\pm^{0.4}_{0.3}\pm$0.5&3.7$\pm$0.7$\pm$0.4&5.0$\pm$0.2$\pm$0.4 \\
$m_{\omega^{\prime\prime}}$, MeV&
1808$\pm^{60}_{40}\pm$20&1760$\pm$50$\pm$40&1750$\pm$20$\pm$6 \\
$\Gamma_{\omega^{\prime\prime}}$, MeV&
807$\pm^{500}_{200}\pm$213&540$\pm^{200}_{100}\pm$50&373$\pm$50$\pm$15 \\
$\sigma(\omega^{\prime\prime}\to 3\pi)$, nb&
1.48$\pm$0.40$\pm$0.46&2.4$\pm$0.7$\pm$0.9&2.7$\pm$0.4$\pm$1.3 \\
$\sigma(\omega^{\prime\prime}\to \omega\pi^+\pi^-)$, nb&
1.54$\pm$0.30$\pm$0.45&1.8$\pm$0.4$\pm$0.5&2.2$\pm$0.3$\pm$0.4 \\
$\sigma(\rho\to 3\pi),$ nb
&&0.13$\pm^{0.06}_{0.04}\pm$0.02& \\
$\phi_{\omega\rho}$, degree  &&-137$\pm^{14}_{10}\pm$7& \\
$\chi^2_\omega/N_{fit}$&(80$\div$120)/49&(56$\div$62)/49&(45$\div$50)/49\\
$\chi^2_{(1)}/N_{fit}$&(21$\div$46)/6&(13.6$\div$16.4)/6&(7.3$\div$11.3)/6\\
$\chi^2_{(SND)}/N_{fit}$&(66$\div$92)/67&(49$\div$56)/67&(58$\div$67)/67\\
$\chi^2_{3\pi(DM2)}/N_{fit}$&(22$\div$37)/18&(22$\div$44)/18&(27$\div$42)/18\\
$\chi^2_{\omega\pi\pi(DM2)}/N$&(11$\div$15)/18&10/18&(23$\div$26)/18\\
$\chi^2_{tot}/N_{fit}$&(182$\div$260)/152&(137$\div$170)/152&(156$\div$180)
/152 \\ \hline
\end{tabular} 
\end{center}
\end{table}

\begin{table}[!hbp]
\caption{Fit results for the $e^+e^-\to\pi^+\pi^-\pi^0$ and $\omega\pi^+\pi^-$
         cross sections. The DM2 data for $e^+e^-\to\pi^+\pi^-\pi^0$ were not
         used. The column number $N$ corresponds to the different
	 models for the $A_{\rho\pi}$ amplitude. $N_{fit}$ is the number of 
         fitted
	 points. The first error is statistical, the second error shows the
	 difference in the fit results due to various assumptions about 
	 the $e^+e^-\to\pi^+\pi^-\pi^0$ reaction dynamics.}
\label{tab5}
\begin{center}
\begin{tabular}[t]{llll}
 $N$&1&2&3 \\ \hline
$m_\omega,$ MeV&782.76$\pm$0.08&782.72$\pm$0.08&782.74$\pm$0.08 \\
$\Gamma_\omega,$ MeV&8.63$\pm$0.04$\pm$0.01&8.71$\pm$0.04$\pm$0.01&8.65$\pm
$0.04 \\
$\sigma(\omega\to 3\pi),$ nb &1608$\pm$7$\pm$1&1618$\pm$10$\pm$2&1612$\pm$7$
\pm$2 \\
$\sigma(\phi\to 3\pi)$, nb&653$\pm$6$\pm$4&652$\pm$8$\pm$5&665$\pm$10$\pm$8 \\
$\phi_{\omega\phi}$, degree &166$\pm$3$\pm$3&165$\pm$3$\pm$4&195$\pm$6$\pm$5 \\
$r_0$, GeV$^{-1}$&&&2.3$\pm^{1.2}_{0.8}\pm$0.2 \\
$m_{\omega^\prime}$, MeV&
1273$\pm^{25}_{20}\pm$28&1386$\pm^{70}_{50}\pm$60&1300$\pm$30$\pm$30 \\
$\Gamma_{\omega^\prime}$, MeV&
405$\pm^{60}_{50}\pm$73&827$\pm^{300}_{200}\pm$186&595$\pm$50$\pm$50 \\
$\sigma(\omega^\prime\to 3\pi)$, nb&
6.9$\pm$0.4$\pm$0.7&5.0$\pm$1.0$\pm$0.5&5.6$\pm$0.3$\pm$0.5 \\
$m_{\omega^{\prime\prime}}$, MeV&
1819$\pm^{90}_{50}\pm$32&1773$\pm^{40}_{30}\pm$12&1758$\pm$20$\pm$5 \\
$\Gamma_{\omega^{\prime\prime}}$, MeV&
679$\pm^{450}_{200}\pm$121&505$\pm^{150}_{100}\pm$35&345$\pm$50$\pm$10 \\
$\sigma(\omega^{\prime\prime}\to 3\pi)$, nb&
5.6$\pm$2.0$\pm$1.1&5.7$\pm$1.7$\pm$0.6&7.6$\pm$1.6$\pm$1.6 \\
$\sigma(\omega^{\prime\prime}\to \omega\pi^+\pi^-)$, nb&
1.2$\pm$0.3$\pm$0.2&1.5$\pm$0.2$\pm$0.1&1.7$\pm$0.2$\pm$0.1 \\
$\sigma(\rho\to 3\pi),$ nb&
&0.083$\pm^{0.056}_{0.033}\pm$0.009& \\
$\phi_{\omega\rho}$, degree  &&-134$\pm^{17}_{13}\pm$8& \\
$\chi^2_\omega/N_{fit}$&(60$\div$63)/49&(52$\div$55)/49&(40$\div$42)/49\\
$\chi^2_{(1)}/N_{fit}$&(10$\div$16.6)/6&(11.9$\div$13)/6&(4.6$\div$6.3)/6\\
$\chi^2_{(SND)}/N_{fit}$&(69$\div$74)/67&(51$\div$52)/67&(52$\div$56)/67\\
$\chi^2_{3\pi(DM2)}/N_{fit}$& -- & -- & -- \\
$\chi^2_{\omega\pi\pi(DM2)}/N$&(11$\div$14)/18&10/18&23/18\\
$\chi^2_{tot}/N_{fit}$&(139$\div$149)/134&(112$\div$118)/134&(115$\div$120)
/134 \\ \hline
\end{tabular} 
\end{center}
\end{table}

\begin{table}[h]
\begin{center}
\caption{Comparison of the $\omega\to 3 \pi$, $\pi^0\gamma$ and $e^+e^-$ decays
branching ratios obtained by using the SND data with the world averages
\cite{pdg}.}
\label{tab6}
\begin{tabular}[t]{ccc}
&SND& PDG-2002  \\ \hline
$B(\omega\to e^+e^-)$    &$(7.52\pm 0.24)\times 10^{-5}$&$(6.95\pm
0.15)\times 10^{-5}$ \\
$B(\omega\to 3\pi)$      &$0.8965\pm 0.0051$&$0.8910 \pm 0.007$ \\
$B(\omega\to\pi^0\gamma)$&$0.0865\pm 0.0045$&$0.087\pm 0.004$ \\
\hline
\end{tabular}
\end{center}
\end{table}

\begin{center}

\begin{figure}
\epsfig{figure=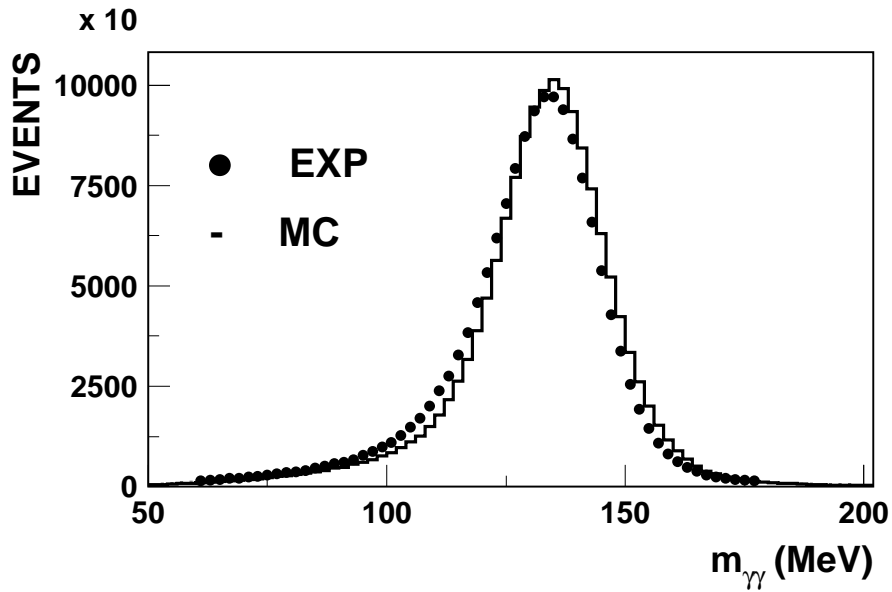,width=13.5cm}
\caption{Two-photon invariant mass distribution in the 
         $e^+e^-\to\pi^+\pi^-\pi^0$ events.}
\label{mppi}
\end{figure}

\begin{figure}
\epsfig{figure=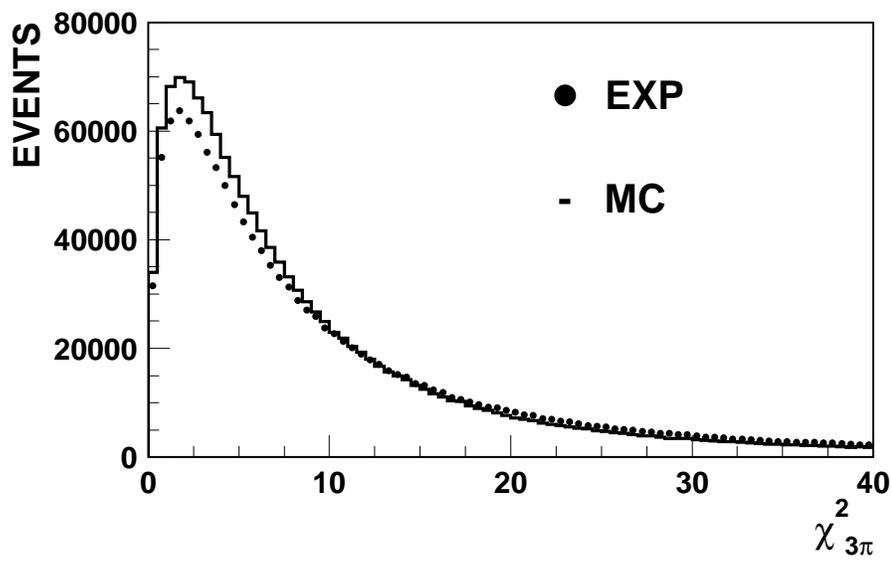,width=13.5cm}
\caption{The $\chi^2_{3\pi}$ distribution in the $e^+e^-\to\pi^+\pi^-\pi^0$ 
events.}
\label{xi2ua}
\end{figure}

\begin{figure}
\epsfig{figure=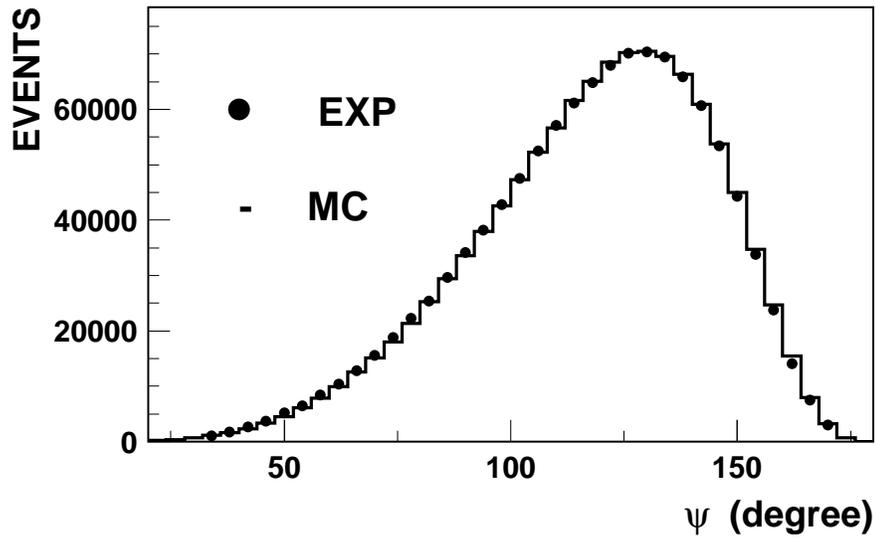,width=13.5cm}
\caption{Angle $\psi$ between the charged pions in the 
$e^+e^-\to\pi^+\pi^-\pi^0$ events.}
\label{an12}
\end{figure}

\begin{figure}
\epsfig{figure=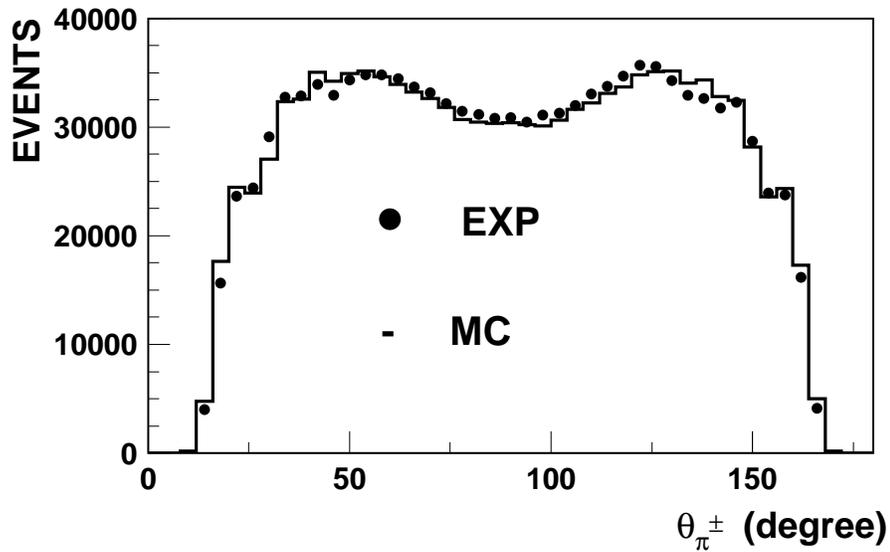,width=13.5cm}
\caption{The $\theta$ distribution of charged pions from the reaction
         $e^+e^- \to \pi^+\pi^-\pi^0$.}
\label{teu12}
\end{figure}

\begin{figure}
\epsfig{figure=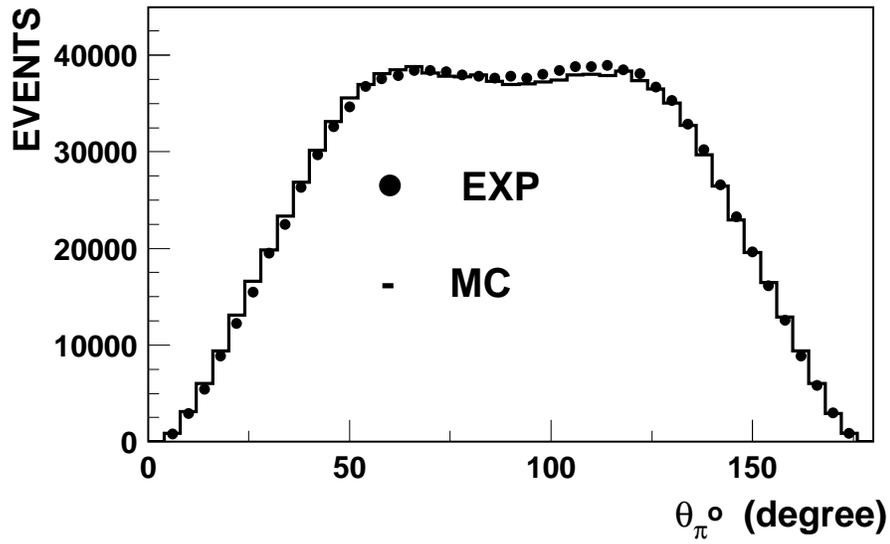,width=13.5cm}
\caption{The $\theta$ distribution of neutral pions from the reaction
         $e^+e^- \to \pi^+\pi^-\pi^0$.}
\label{teu3}
\end{figure}

\begin{figure}
\epsfig{figure=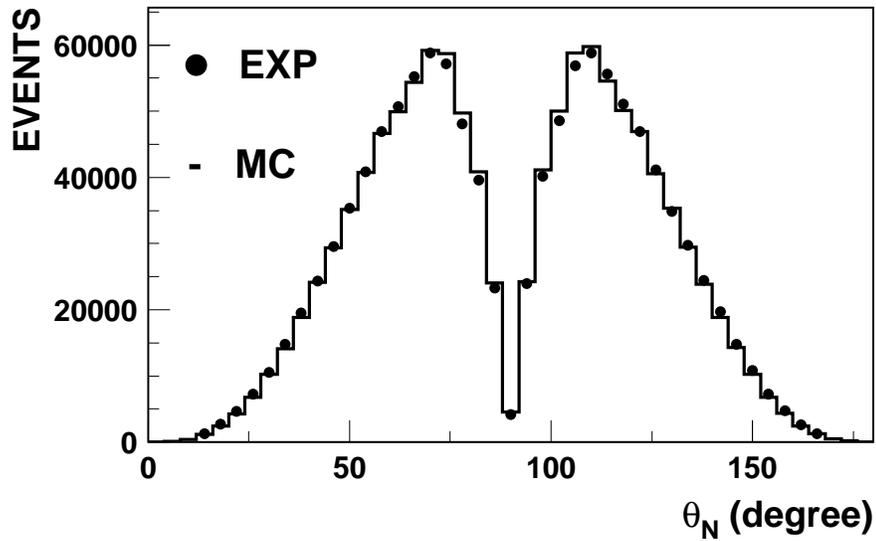,width=13.5cm}
\caption{The angle between the normal to the production plane and $e^+e^-$
         beam direction for the $e^+e^- \to \pi^+\pi^-\pi^0$ events.}
\label{sohe}
\end{figure}

\begin{figure}
\epsfig{figure=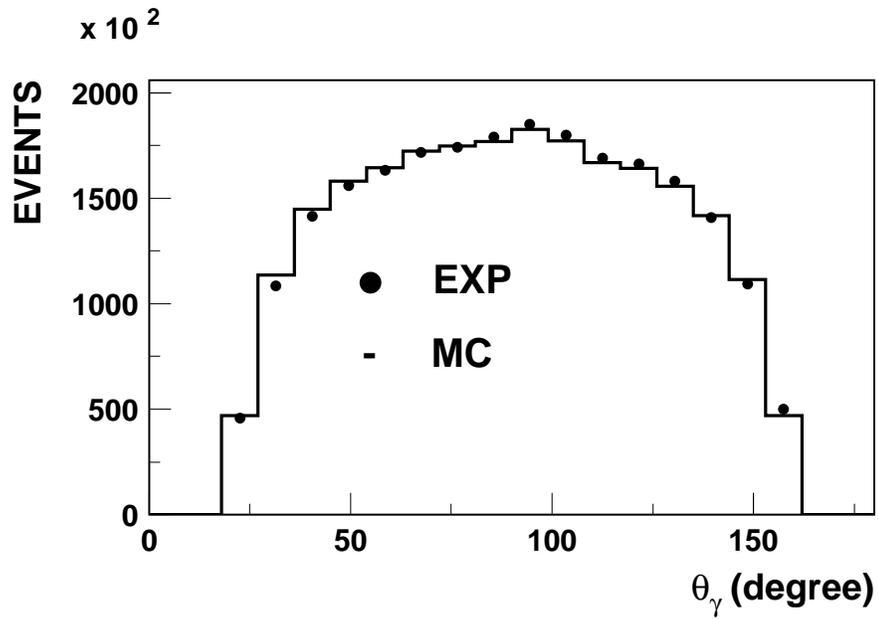,width=13.5cm}
\caption{The angular distribution of photons.}
\label{teu45}
\end{figure}

\begin{figure}
\epsfig{figure=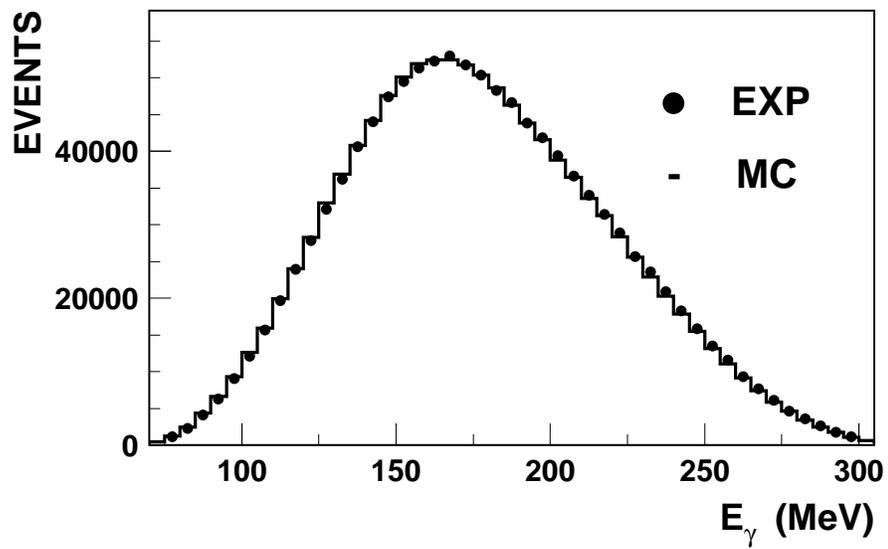,width=13.5cm}
\caption{The energy distribution for the most energetic photon.}
\label{epu4}
\end{figure}

\begin{figure}
\epsfig{figure=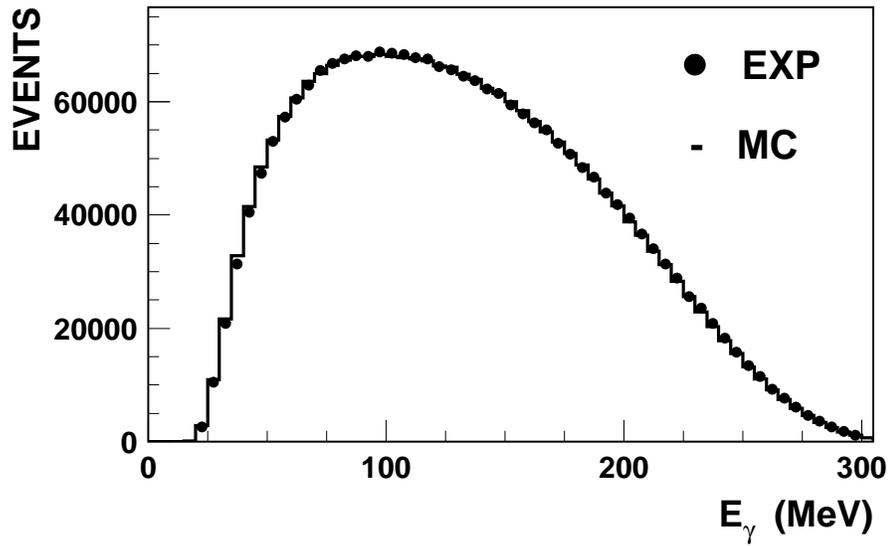,width=13.5cm}
\caption{The photon energy distribution.}
\label{epu45}
\end{figure}

\begin{figure}
\epsfig{figure=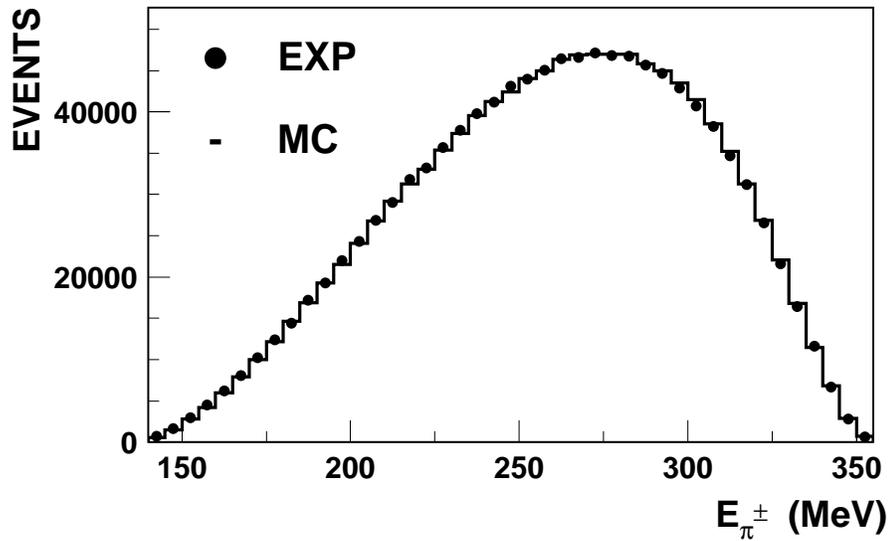,width=13.5cm}
\caption{The charged pions energy distribution}
\label{epu12}
\end{figure}

\begin{figure}
\epsfig{figure=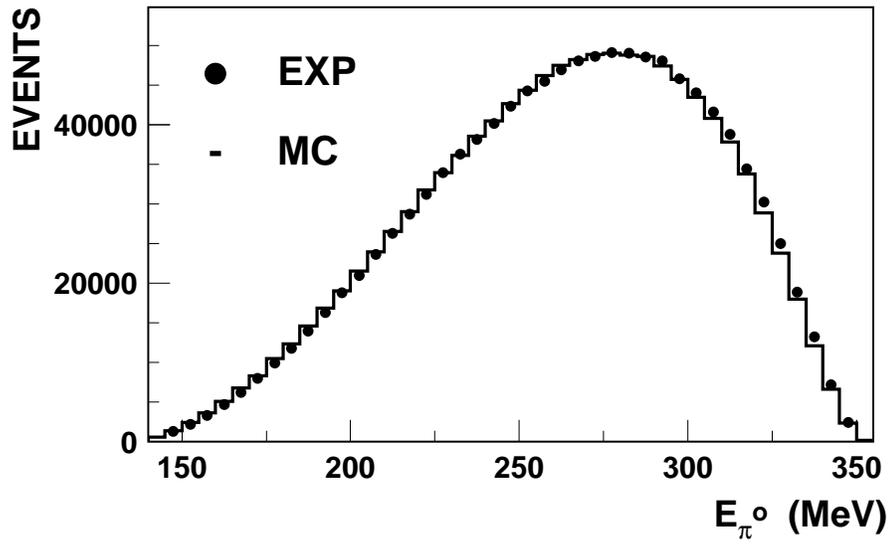,width=13.5cm}
\caption{The neutral pions energy distribution}
\label{epu3}
\end{figure}

\begin{figure}
\epsfig{figure=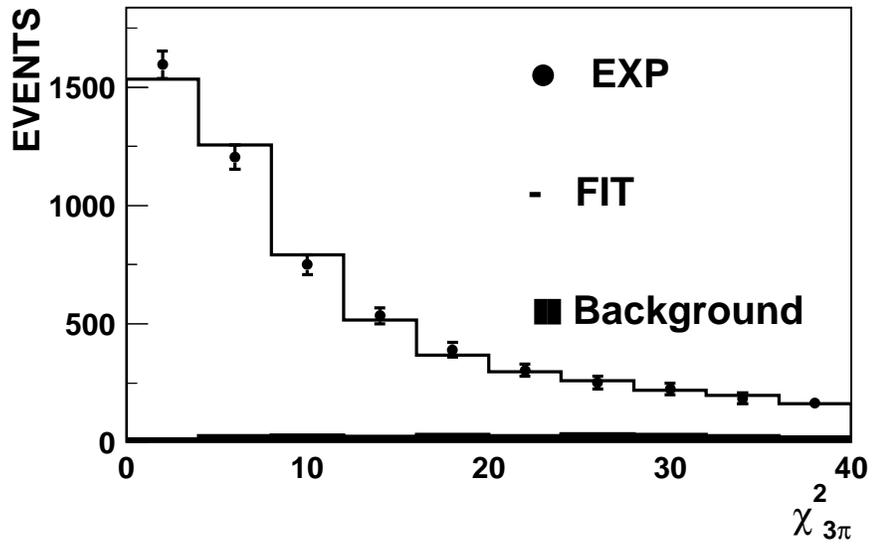,width=13.5cm}
\caption{The experimental $\chi^2_{3\pi}$ distribution in the energy region
         $\sqrt[]{s}>870$ MeV, fitted by a sum of distributions for the 
         signal and background. 
         The background contribution is shown by the filled histogram.}
\label{xi2ufit}
\end{figure}

\begin{figure}
\epsfig{figure=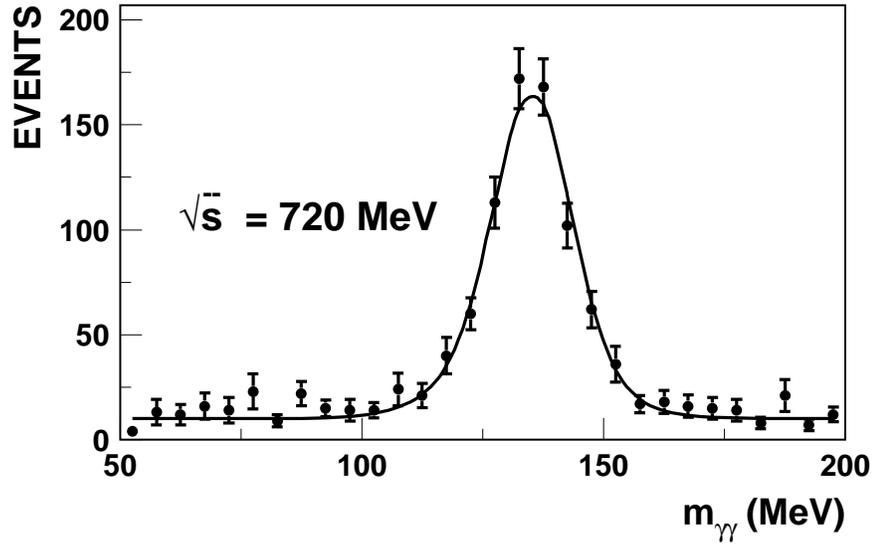,width=13.5cm}
\caption{The two-photon invariant mass distribution at $\sqrt[]{s}=720$ MeV.}
\label{mpgg360}
\end{figure}

\begin{figure}
\epsfig{figure=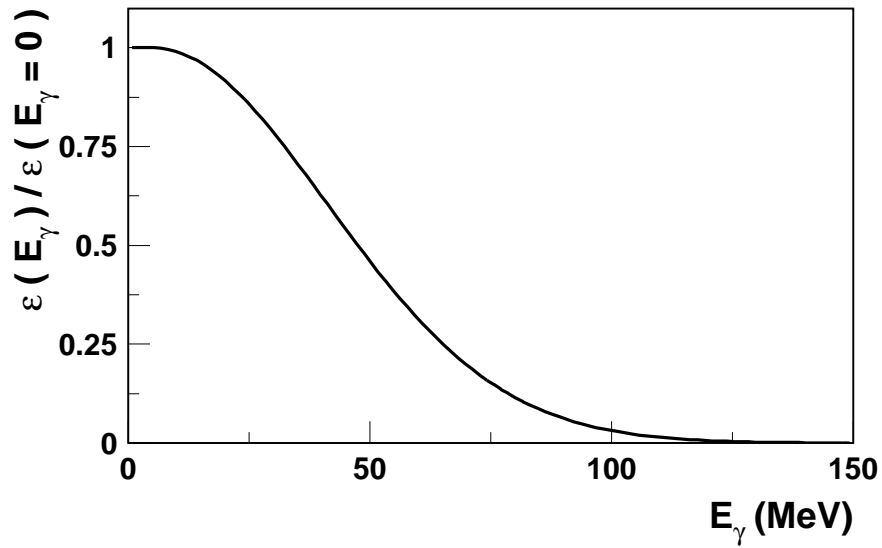,width=13.5cm}
\caption{The detection efficiency $\epsilon(E_\gamma)$ dependence on the
radiated photon energy $E_\gamma$ for the $e^+e^-\to\pi^+\pi^-\pi^0(\gamma)$
events at $\sqrt[]{s}\simeq m_\omega$, obtained by simulation.}
\label{efrad}
\end{figure}

\begin{figure}
\epsfig{figure=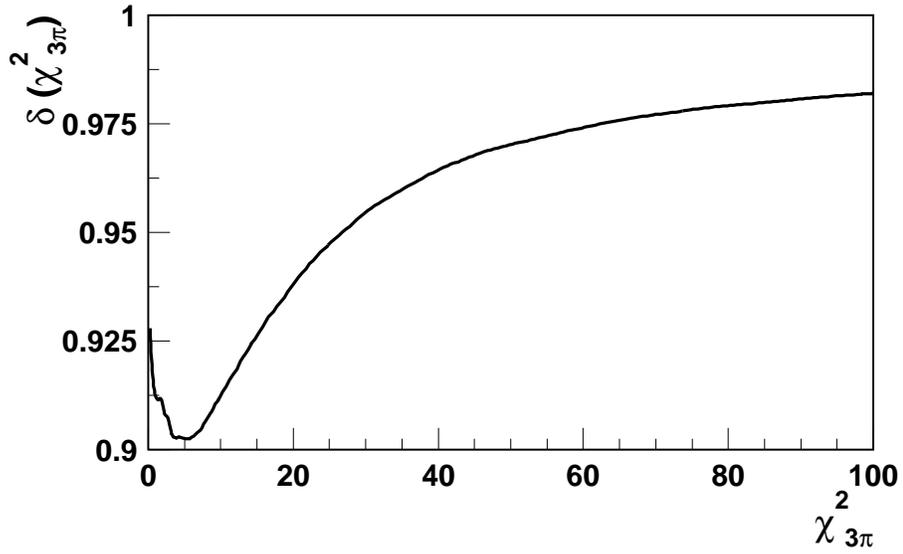,width=13.5cm}
\caption{The correction coefficient $\delta_{\chi^2_{3\pi}}$ dependence on
the value of the cut on $\chi^2_{3\pi}$.}
\label{xi2ud}
\end{figure}

\begin{figure}
\epsfig{figure=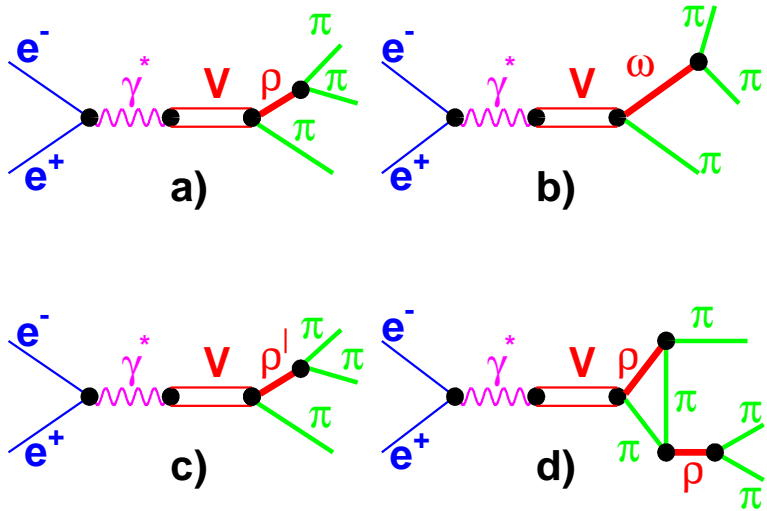,width=13.5cm}
\caption{The $e^+e^-\to \pi^+\pi^-\pi^0$ transition diagrams.}
\label{dplfeiman}
\end{figure}

\begin{figure}
\epsfig{figure=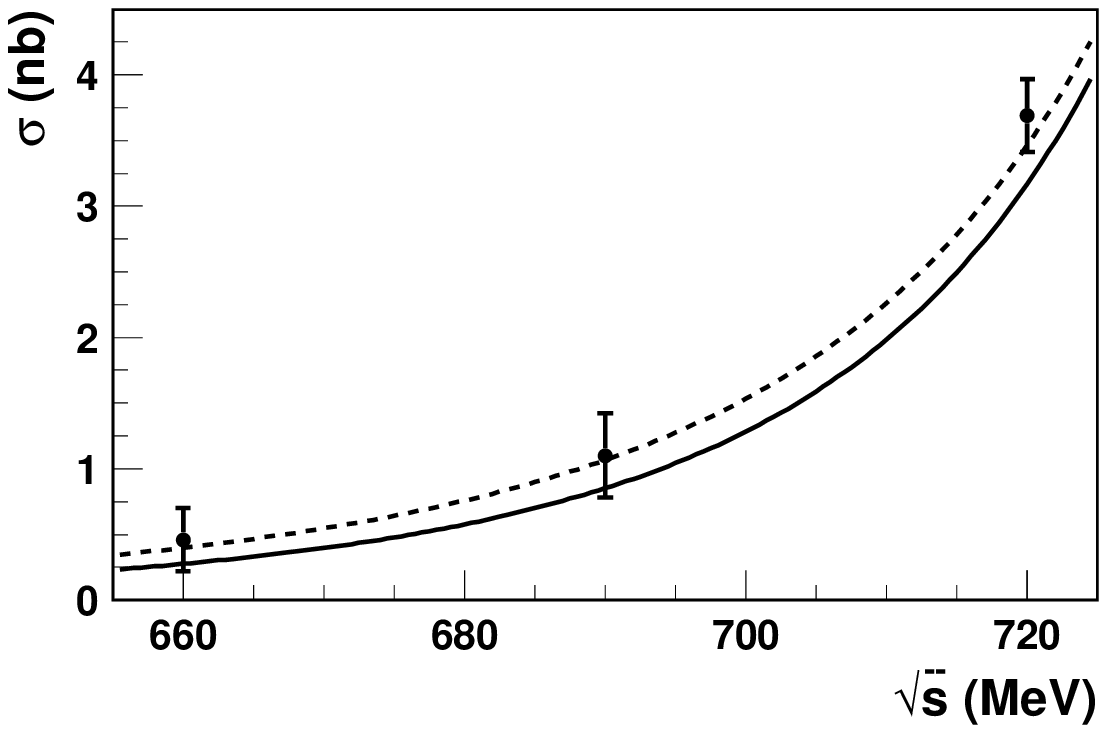,width=12.5cm}
\caption{The $e^+e^-\to\pi^+\pi^-\pi^0$ cross section. Dots are the SND
data obtained in this work. Curves are results of fitting to the data in
the model 2 (solid curve) and in the model 3 (dashed curve).}
\label{crf1}
\end{figure}

\begin{figure}
\epsfig{figure=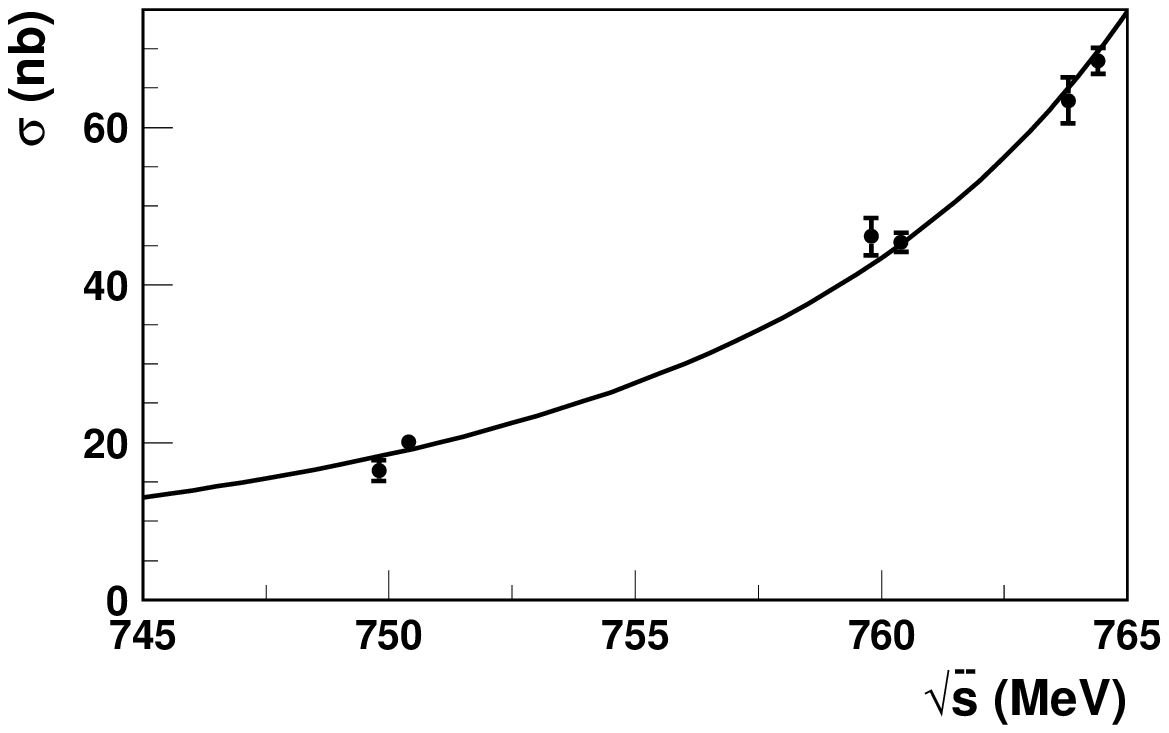,width=12.5cm}
\caption{The $e^+e^-\to\pi^+\pi^-\pi^0$ cross section. Dots are the SND
data obtained in this work; the curve is the fit result.}
\label{crf2}
\end{figure}

\begin{figure}
\epsfig{figure=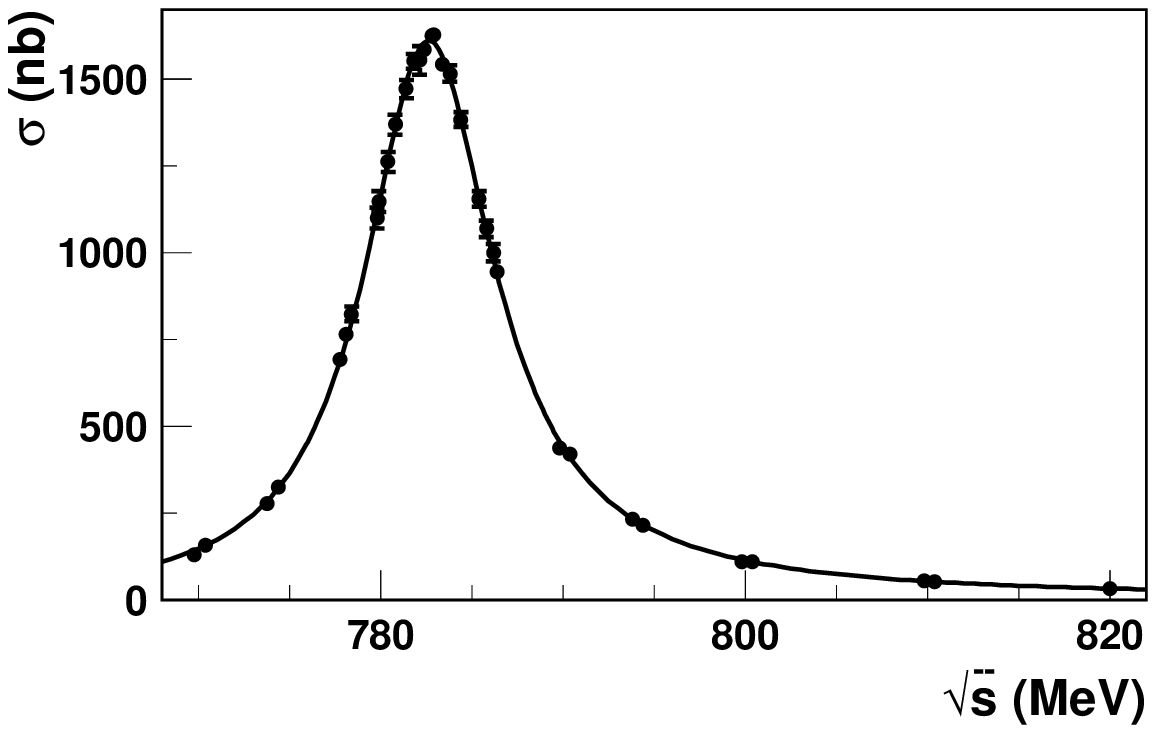,width=12.5cm}
\caption{The $e^+e^-\to\pi^+\pi^-\pi^0$ cross section. Dots are the SND
data obtained in this work; the curve is the fit result.}
\label{crf3}
\end{figure}

\begin{figure}
\epsfig{figure=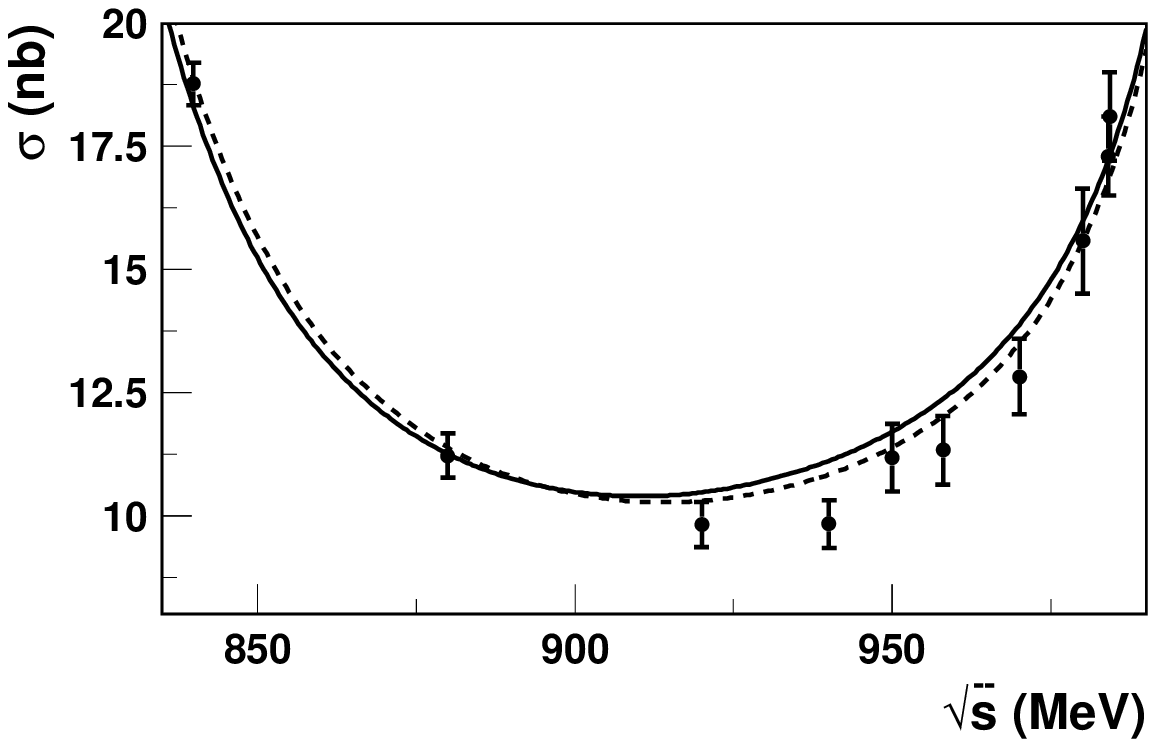,width=12.5cm}
\caption{The $e^+e^-\to\pi^+\pi^-\pi^0$ cross section. Dots are the SND
data obtained in this work and in Ref.\cite{phi98,pi3mhad}. Curves are results
of fitting to the data in the model 2 (solid curve) and in the model 3 
(dashed curve).}
\label{crf4}
\end{figure}

\begin{figure}
\epsfig{figure=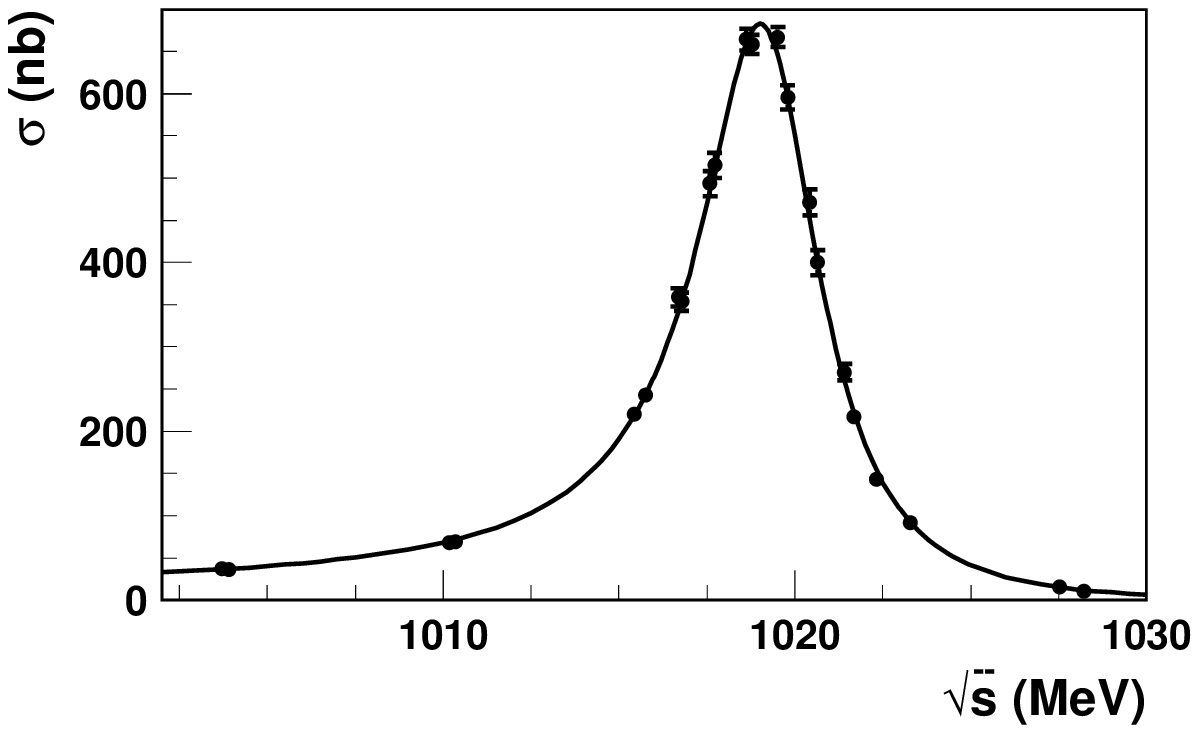,width=12.5cm}
\caption{The $e^+e^-\to\pi^+\pi^-\pi^0$ cross section.  Dots are the SND
data obtained in Ref.\cite{phi98}; the curve is the fit result.}
\label{crf5}
\end{figure}

\begin{figure}
\epsfig{figure=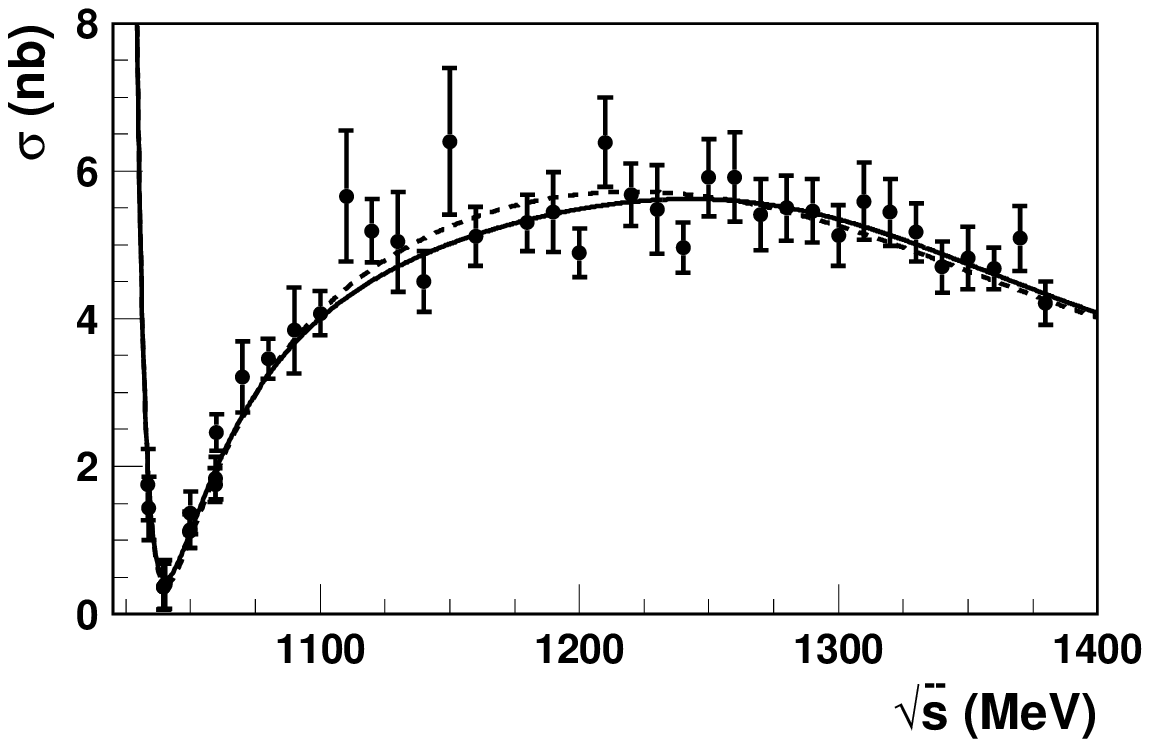,width=12.5cm}
\caption{The $e^+e^-\to\pi^+\pi^-\pi^0$ cross section. Dots are the 
experimental
data obtained in Ref.\cite{phi98,pi3mhad}. Curves are results of fitting to
the data in the model 2 (solid curve) and in the model 3 (dashed curve).}
\label{crf6}
\end{figure}

\begin{figure}
\epsfig{figure=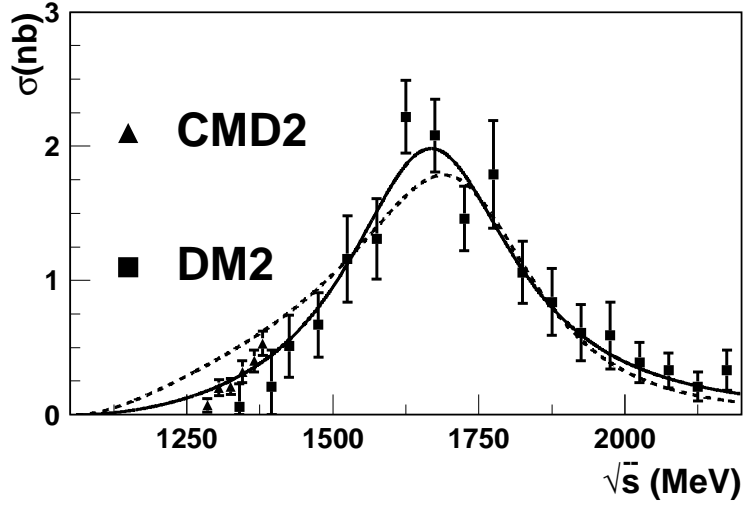,width=12.5cm}
\caption{The $e^+e^-\to\omega\pi^+\pi^-$ cross section. The results of the DM2
\cite{dm2} and CMD2 \cite{kmd2opp} experiments are shown. Curves are results
of fitting to the DM2 data in the model 2 (solid curve) and in the model 3 
(dashed curve).}
\label{crsopp}
\end{figure}

\begin{figure}
\epsfig{figure=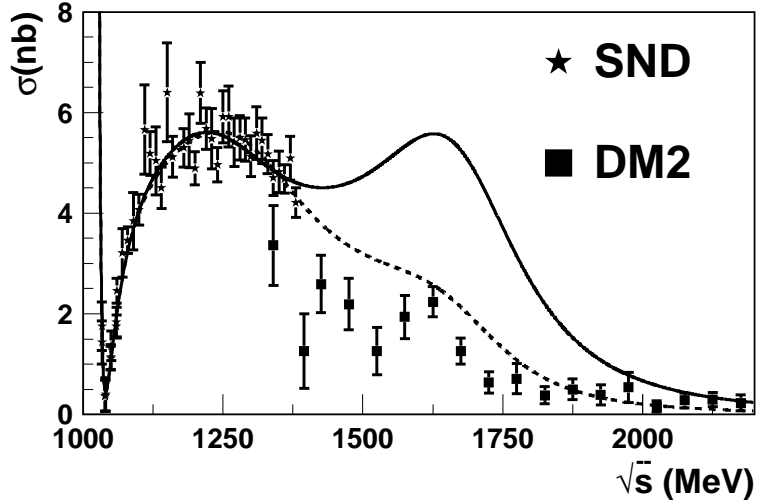,width=12.5cm}
\caption{The $e^+e^-\to\pi^+\pi^-\pi^0$ cross section. The results of the SND
\cite{phi98,pi3mhad} and DM2 \cite{dm2} are shown.Curves are results of
fitting to the data in the model 2. Dashed curve corresponds to the fit
under assumption that a relative bias between the SND and DM2 data exists
(DM2 data were scaled by a factor of $C=1.54$). Solid curve is the result of 
the fitting to the SND data only.}
\label{crf8}
\end{figure}

\begin{figure}
\epsfig{figure=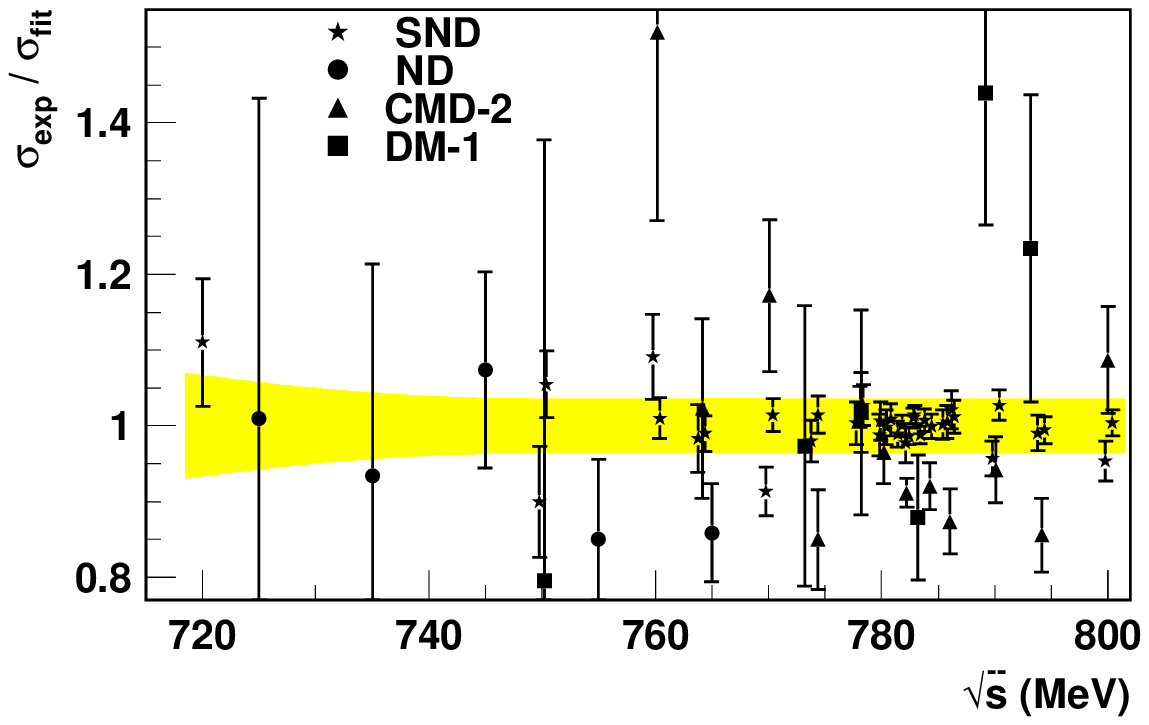,width=12.5cm}
\caption{The ratio of the $e^+e^-\to\pi^+\pi^-\pi^0$ cross section obtained
in different experiments to the fit curve. The shaded area shows the
systematic error of the SND measurements. The SND (this work), 
DM1 \cite{cord}, ND
\cite{nd,ndnerez}, CMD2 \cite{kmd2omega} results are presented.}
\label{raz1}
\end{figure}

\begin{figure}
\epsfig{figure=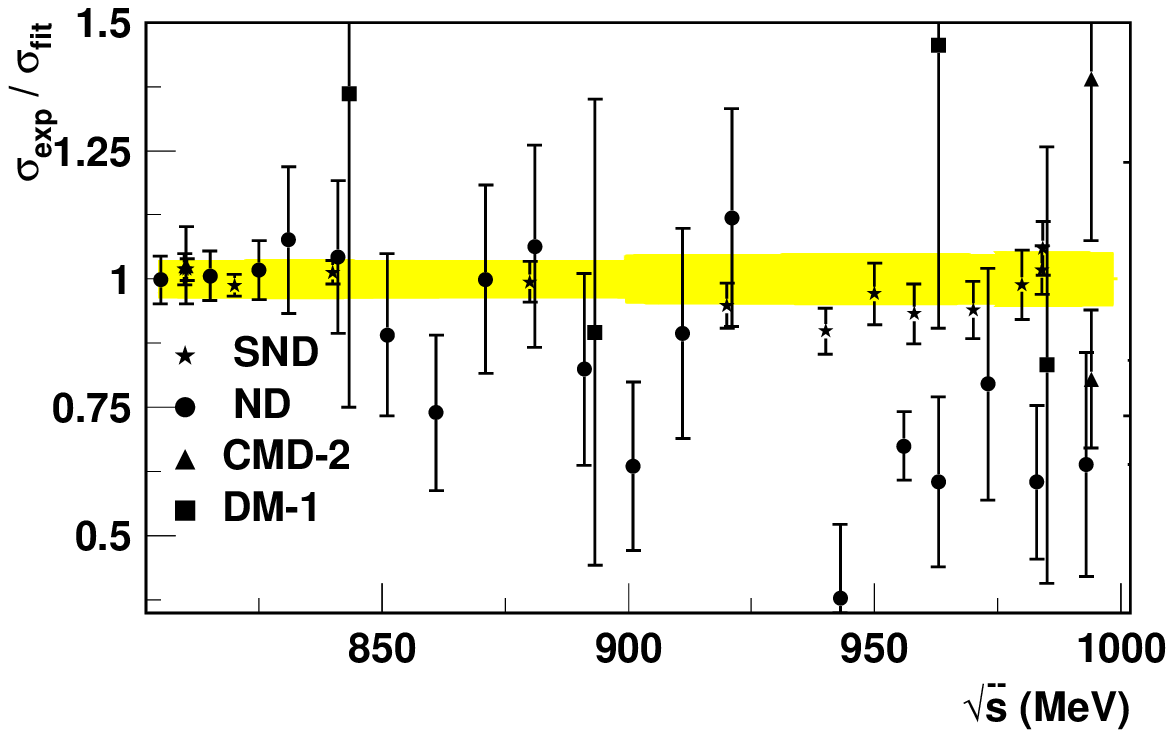,width=12.5cm}
\caption{The ratio of the $e^+e^-\to\pi^+\pi^-\pi^0$ cross section obtained
in different experiments to the fit curve. The shaded area shows the
systematic error of the SND measurements. The SND (this work and
Ref.\cite{phi98,pi3mhad}) DM1 \cite{cord}, ND \cite{nd,ndnerez}, CMD2
\cite{kmd2phi1,kmd2phi2} results are presented.}
\label{raz2}
\end{figure}

\begin{figure}
\epsfig{figure=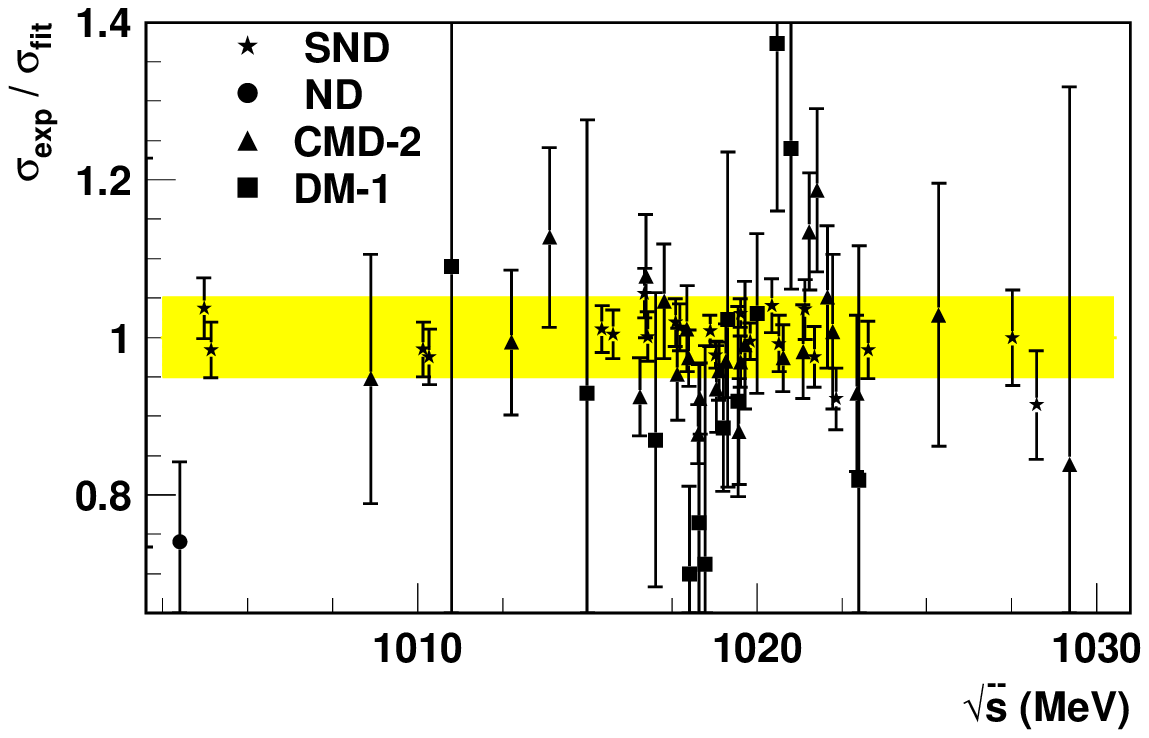,width=12.5cm}
\caption{The ratio of the $e^+e^-\to\pi^+\pi^-\pi^0$ cross section obtained
in different experiments to the fit curve. The shaded area shows the
systematic error of the SND measurements. The SND \cite{phi98}, 
DM1 \cite{cord},
ND \cite{nd,ndnerez}, CMD2 \cite{kmd2phi1,kmd2phi2} results are presented.}
\label{raz3}
\end{figure}

\begin{figure}
\epsfig{figure=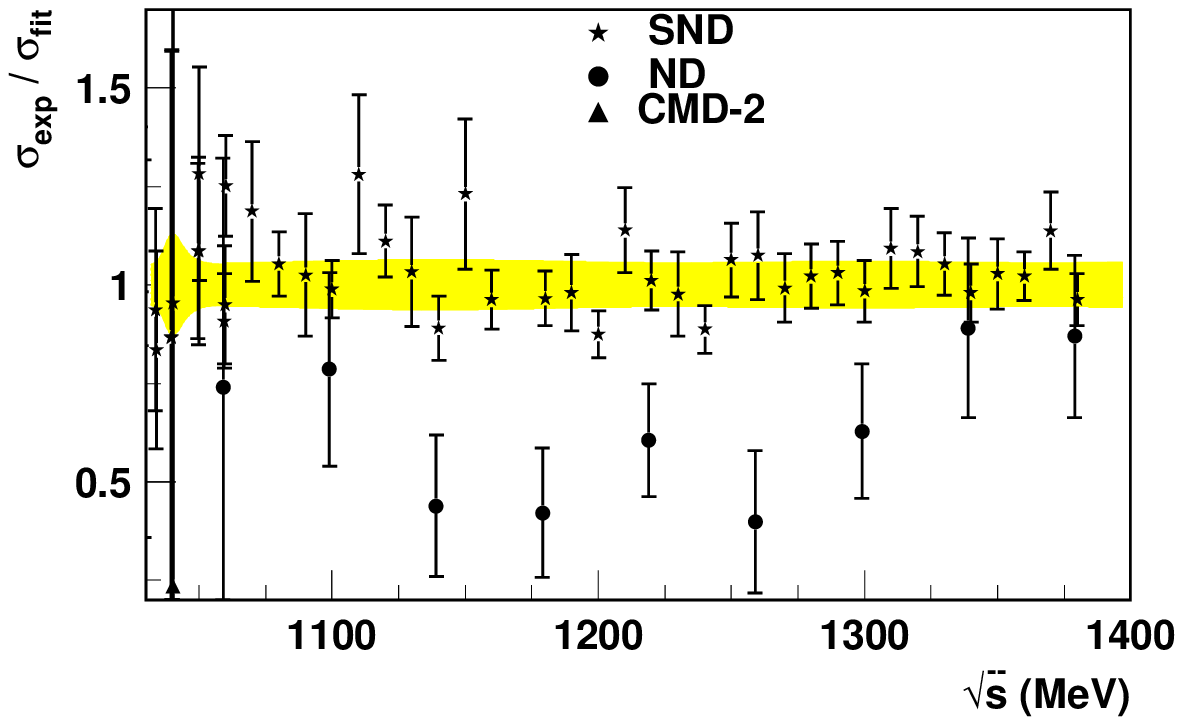,width=12.5cm}
\caption{The ratio of the $e^+e^-\to\pi^+\pi^-\pi^0$ cross section obtained in
different experiments to the fit curve. The shaded area shows the
systematic error of the SND measurements. The SND \cite{phi98,pi3mhad}, ND
\cite{nd,ndnerez}, CMD2 \cite{kmd2phi2} results are presented.}
\label{raz4}
\end{figure}

\begin{figure}
\epsfig{figure=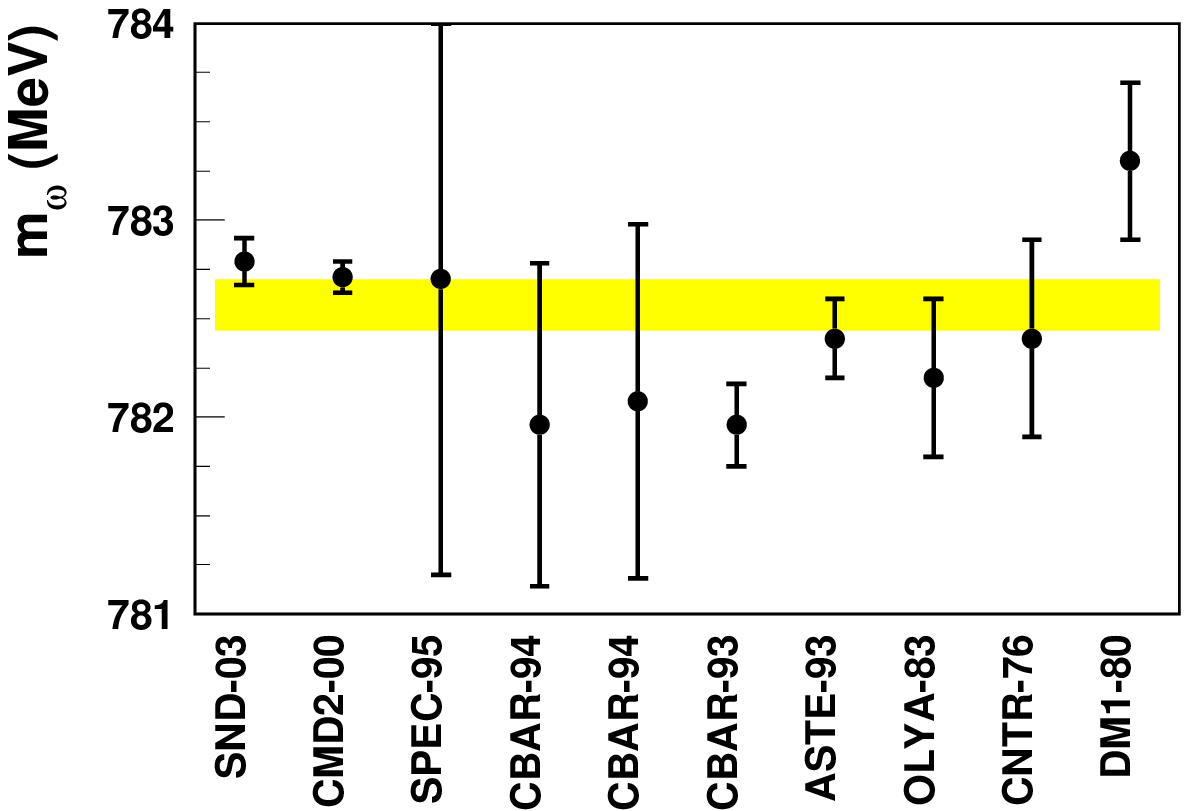,width=12.5cm}
\caption{The $\omega$ meson mass $m_\omega$ measured in this work (SND-03) and
in Ref.\cite{kmd2omega,spec95,cbar,olyaomega,cntr76,cord}.
The shaded area shows the world average value \cite{pdg}.}
\label{massa}
\end{figure}

\begin{figure}
\epsfig{figure=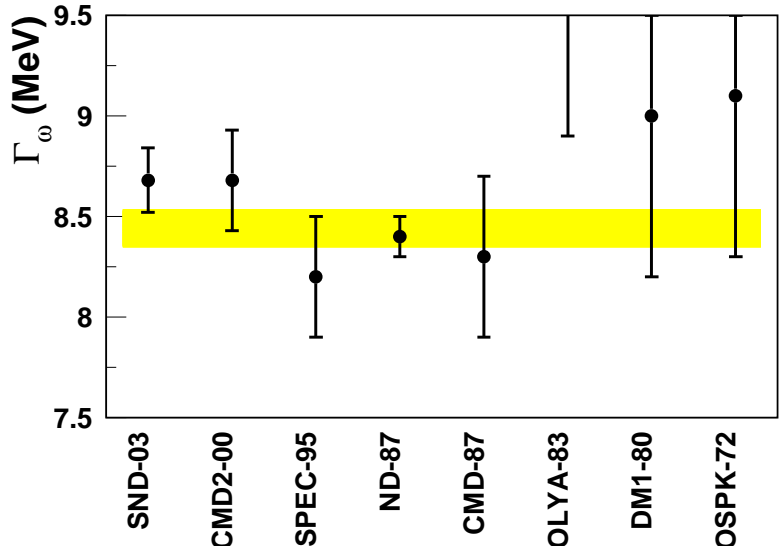,width=12.5cm}
\caption{The $\omega$ meson width $\Gamma_\omega$ measured in this work
(SND-03) and in Ref.\cite{kmd2omega,spec95,ndomega,kmd,olyaomega,cord,benak}.
The shaded area shows the world average value \cite{pdg}.}
\label{shira}
\end{figure}

\begin{figure}
\epsfig{figure=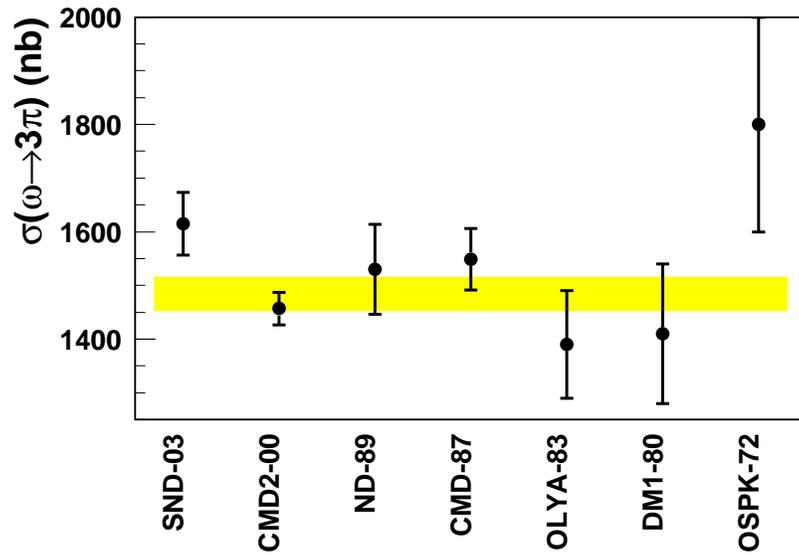,width=12.5cm}
\caption{The value of $\sigma(\omega\to 3\pi)$ measured in this work (SND-03)
and in Ref.\cite{kmd2omega,ndomega,kmd,olyaomega,cord,benak}.
The shaded area shows the world average value \cite{pdg}.}
\label{sech}
\end{figure}

\begin{figure}
\epsfig{figure=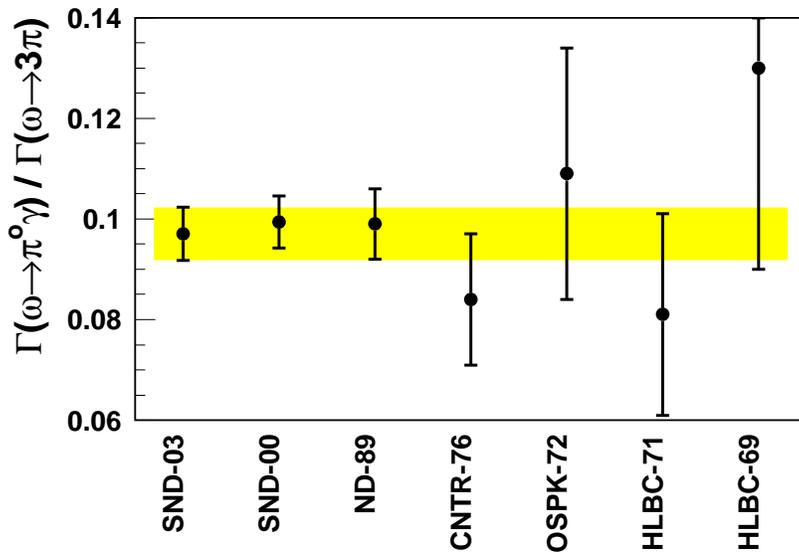,width=12.5cm}
\caption{The ratio of the partial widths $\omega\to\pi^0\gamma$ and
$\omega\to 3\pi$, obtained in this work (SND-03) and in
Ref.\cite{sndot,ndomega,cntr76,benak,hlbc}. 
The shaded area shows the world average value \cite{pdg}.}
\label{otnshir}
\end{figure}

\begin{figure}
\epsfig{figure=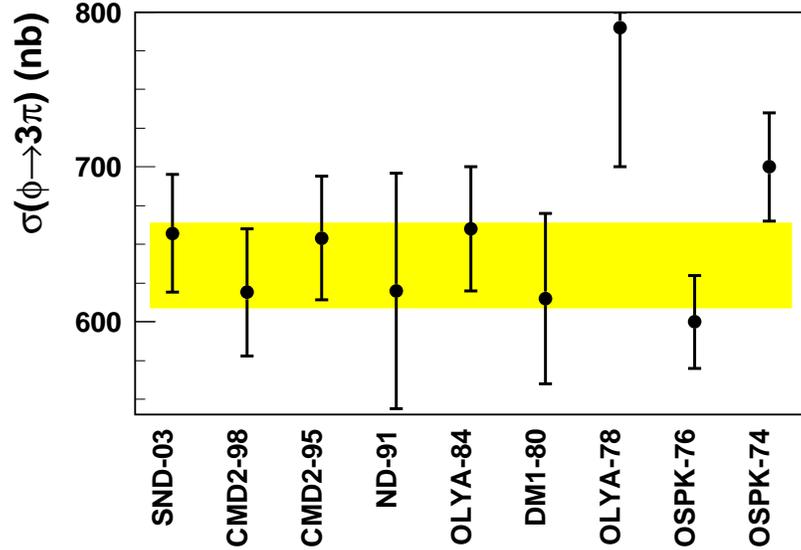,width=12.5cm}
\caption{The value of $\sigma(\phi\to 3\pi)$ obtained in this work (SND-03) and
in Ref.\cite{kmd2phi2,kmd2phi1,nd,olyaphiomega,cord,buk,parr,cosm}.
The shaded area shows the world average value according to the year 
2000 PDG table \cite{pdg2000}.}
\label{phisech}
\end{figure}

\begin{figure}
\epsfig{figure=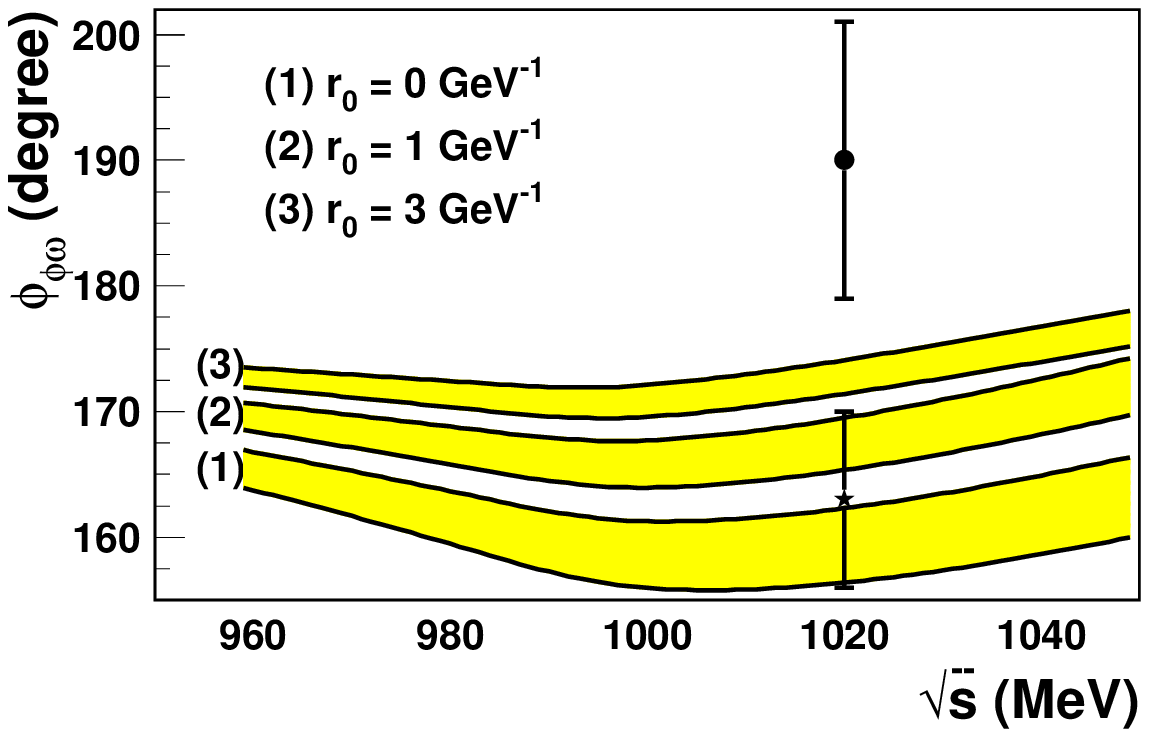,width=12.5cm}
\caption{The phase $\phi_{\omega\phi}$ obtained in this work. The star and 
the dot indicate
the phase values obtained from the fit in the second and third models
respectively. The shaded areas show the expected
energy behavior of the $\phi_{\omega\phi}$ phase \cite{faza} for various values
of the range parameter $r_0$.}
\label{faza}
\end{figure}

\end{center}


\begin{thebibliography}{99}
\bibitem{pdg}
 K. Hagivara et al., Phys. Rev. {\bf D 66}, 010001 (2002)
\bibitem{pi3mhad}
 M.N. Achasov et al., Phys. Rev. {\bf D 66}, 032001 (2002)
\bibitem{thrhoom}
 N.N. Achasov, A.A. Kozhevnikov, and G.N. Shestakov, Phys. Lett. {\bf 50B},
 448 (1974) .
 N.N. Achasov, N.M. Budnev, A.A. Kozhevnikov, and G.N. Shestakov,
 Yad. Fiz. 23, 610 (1976) [Sov. J. Nucl. Phys. 23, 320 (1976)];
 N.N. Achasov and G.N. Shestakov, Fiz. Elem. Chastits. At. Yadra 9, 48 (1978) 
\bibitem{aug}
 J.E. Augustin et al., Phys. Lett. {\bf B 28}, 513 (1969)
\bibitem{benak}
 D. Benaksas et al., Phys. Lett. {\bf B 42}, 507 (1972)
\bibitem{olyaomega}
 L.M. Kurdadze, et al, JETP Lett. 36, 274 (1982) 
 [Pisma Zh. Eksp. Teor. Fiz. 36, 221 (1982)]
\bibitem{ndomega}
 V.M. Aulchenko et al., Phys. Lett. {\bf B 186}, 432 (1987); 
 S.I. Dolinsky et al., Z. Phys. {\bf C 42}, 511 (1989)
\bibitem{nd}
 S.I. Dolinsky et al., Phys. Rep. {\bf 202}, 99 (1991)
\bibitem{kmd}
 L.M. Barkov et al., JETP Lett. 46, 164 (1987)
 [Pisma Zh. Eksp. Teor. Fiz. 46, 132 (1987)]
\bibitem{kmd2omega}
 R.R. Akhmetshin et al., Phys. Lett. {\bf B 476}, 33 (2000)
\bibitem{cosm}
 G. Cosme et al., Phys. Lett. {\bf B 48}, 155 (1974)
\bibitem{parr}
 G.Parrour et al., Phys. Lett. {\bf B 63}, 357 (1976)
\bibitem{parr2}
 G.Parrour et al., Phys. Lett. {\bf B 63}, 362 (1976)
\bibitem{buk}
 A.D. Bukin et al., Yad. Fiz. 27, 976 (1978) 
 [Sov. J. Nucl. Phys. 27, 516 (1978)]
\bibitem{kmd2phi1}
 R.R. Akhmetshin et al., Phys. Lett. {\bf B 364}, 199 (1995)
\bibitem{kmd2phi2}
 R.R. Akhmetshin et al., Phys. Lett. {\bf B 434}, 426 (1998)
\bibitem{cord}
 A. Cordier et al., Nucl. Phys. {\bf B 172}, 13 (1980)
\bibitem{olyaphiomega}
 L.M. Kurdadze et al., Preprint, Budker INP 84-7, Novosibirsk, 1984
 (in Russian)
\bibitem{ndnerez}
 A.D. Bukin et al., Yad. Fiz. 50, 999 (1989)
 [Sov. J. Nucl. Phys. 50, 621 (1989)]
\bibitem{ndrho}
 I.B. Vasserman et al., Yad. Fiz. 48, 753 (1988)
 [Sov. J. Nucl. Phys. 48, 480 (1988)]
\bibitem{m3n}
 G. Cosme et al., Nuc. Phys. {\bf B 152}, 215 (1979)
\bibitem{mea}
 B. Esposito et al., Lett. Nuovo Cim. 28, 195 (1980)
\bibitem{gg2}
 C. Bacci et al., Nuc. Phys. {\bf B 184}, 31 (1981)
\bibitem{dm1}
 B. Delcourt et al., Phys. Lett {\bf 113B}, 93 (1982)
\bibitem{dm2}
 A. Antonelli et al., Z. Phys., {\bf C 56}, 15 (1992)
\bibitem{sndmhad}
 M.N. Achasov et al., Phys. Lett. {\bf B 462}, 265 (1999)
\bibitem{phi98}
 M.N. Achasov et al., Phys. Rev. {\bf D 63}, 072002 (2001)
\bibitem{dplphi98}
 M.N. Achasov et al., Phys. Rev. {\bf D 65}, 032002 (2002)
\bibitem{sndnim}
 M.N. Achasov et al., Nucl. Instr. and Meth. {\bf A 449}, 125 (2000)
\bibitem{vepp2}
 A.N. Skrinsky, in Proc. of Workshop on physics and detectors for
 DA$\Phi$NE, Frascati, Italy, April 4-7, 1995, p.3 
\bibitem{rdp}
 A.N. Skrinsky and Yu.M. Shatunov, Sov. Phys. Uspekhi 32 (1989) 548
\bibitem{ppg}
 M.N.Achasov et al., Phys. Lett. {\bf B 537}, 201 (2002)
\bibitem{akfaz}
 N.N. Achasov and A.A.  Kozhevnikov, Phys. Rev. D 49, 5773 (1994) 
 Yad. Fiz. 56, 191 (1993) [ Phys. Atom. Nucl. 56, 1261 (1993)] 
 Int. J. Mod. Phys. A 9, 527 (1994)
\bibitem{akozi}
 N.N. Achasov and A.A. Kozhevnikov, Yad. Fiz. 55, 809 (1992)
 [Sov. J. Nucl. Phys. 55, 449 (1992)];
 Int. J. Mod. Phys.  A 7, 4825 (1992).
\bibitem{ak2}
 N.N. Achasov, A.A. Kozhevnikov, Phys. Rev. {\bf D 57}, 
 4334 (1998) 
 N.N. Achasov, A.A. Kozhevnikov, Yad. Fiz. {\bf 60}, 
 2212 (1997) [Phys. At. Nucl. {\bf 60}, 2029 (1997)]  
\bibitem{freshlook}
 N.N. Achasov et al., Yad. Fiz. {\bf 54}, 1097 (1991)
 [Sov. J. Nucl. Phys. {\bf 54}, 664 (1991)];
 N.N. Achasov, et. al., Int. J. of Mod. Phys. {\bf A vol.7 No.14}
 3187 (1992)
\bibitem{pi0gam} 
 M.N. Achasov et al., Phys. Lett. {\bf B 559}, 171 (2003)
\bibitem{etagam}
 M.N. Achasov et al., Zh. Eksp. Teor. Fiz. {\bf 117}, 22 (2000)
 [JETP {\bf 90}, 17 (2000)]
\bibitem{faza}
 N.N. Achasov, A.A. Kozhevnikov, Phys. Rev. {\bf D 61} 054005 (2000) 
 Yad. Fiz. 63, 2029 (2000) [ Phys. Atom. Nucl. 63, 1936 (2000)] 
\bibitem{fadin}
 E.A. Kuraev, V.S. Fadin, Yad. Fiz. {\bf 41}, 733 (1985)
 [Sov. J. Nucl. Phys. {\bf 41}, 466 (1985)] 
\bibitem{kmd2opp}
 R.R.Akhmetshin et al., Phys. Lett. {\bf B 489}, 125 (2000)
\bibitem{spec95}
 R. Wurzinger et al., Phys. Rev. {\bf C 51} 443 (1995)
\bibitem{cbar}
 C. Amsler et al., Phys. Lett. {\bf B 327} 425 (1994) 
 C. Amsler et al., Phys. Lett. {\bf B 311} 362 (1993) 
 P. Weidenauer et al., Z. Phys {\bf C 59} 387 (1993)
\bibitem{cntr76}
 J. Keyne et al., Phys. Rev. {\bf D 14} 28 (1976)
\bibitem{sndot}
 V.M. Aulchenko et al., Zh. Eksp. Teor. Fiz. {\bf 117}, 1067 (2000) 
 [V.M. Aulchenko et al., JETP 90, 927 (2000)]
\bibitem{hlbc}
 A.B. Baldin et al., Yad. Fiz. 13, 1318 (1971)
 [Sov. J. Nucl. Phys. 13, 758 (1971)] 
 F. Jacquet et al., Nuovo Cim. {\bf A 63} 743 (1969)
\bibitem{pionuc}
 T.Ericson and W.Wise, Pions and Nuclei, Clarendon Press, Oxford, 1988.
\bibitem{pdg2000}
 D.E. Groom et al., Eur. Phys. J. {\bf C 15}, 1 (2000).
\bibitem{akphiom1}
 N.N. Achasov,  A.A. Kozhevnikov, Phys. Lett. {\bf B 233}, 474 (1989)
\bibitem{akphiom2}
 N.N. Achasov,  A.A. Kozhevnikov, Yad. Fiz. {\bf 55}, 3086 (1992)
 [ Sov. J. Nucl. Phys. 55, 1726 (1992)] 
 Part. World {\bf 3}, 125 (1993)
\bibitem{akphiom3}
 N.N. Achasov, A.A. Kozhevnikov, Phys. Rev. {\bf D 52} 3119 (1995);
 Yad. Fiz. {\bf 59}, 153 (1996) [Phys. Atom. Nucl., {\bf 59}, 144 (1996)]
\bibitem{sndmumu}
 M.N. Achasov et al., Phys. Lett {\bf B 456}, 304 (1999) 
 Phys. Rev. Lett. {\bf 86}, 1698 (2001)
\bibitem{shiree}
 Van Royan, V.F. Weisskopf, Nuovo Cim {\bf A 50}, 617 (1967)
\bibitem{arbuz}
 A.B. Arbuzov et al., JHEP 9710:001 (1997) 
\bibitem{aki}
 N.N. Achasov, A.V.Kiselev, Pisma Zh. Eksp. Teor. Fiz. {\bf75}, 643 (2002)
 [JETP Lett. {\bf 75}, 527 (2002)] 
\bibitem{vepp2000}
 Yu.M.Shatunov et al, Project of a new electron-positron collider VEPP-2000,
 in Proc. of the 2000 European Particle Acc. Conf., Vienna (2000), p.439 

\end{thebibliography}
\end{document}